\documentclass[aps,10pt,eqsecnum,preprint,nofootinbib,superscriptaddress]{revtex4}
\usepackage[utf8]{inputenc}
\usepackage{amssymb,amsmath,amsthm,graphicx,amscd,mathtools}
\usepackage{upgreek,enumerate,color,comment,ulem,slashed,stackrel}
\usepackage{bm,pbox,cancel}
\usepackage[hidelinks]{hyperref}
\usepackage{orcidlink}

\begin{document}

\title{Entanglement dynamics of coupled quantum oscillators in independent nonMarkovian baths}
\author{Jen-Tsung Hsiang\orcidlink{0000-0002-9801-208X}}
\email{cosmology@gmail.com}
\affiliation{Center for High Energy and High Field Physics, National Central University, Taoyuan 320317, Taiwan, ROC}
\author{Onat Ar{\i}soy}
\email{oarisoy@umd.edu}
\affiliation{Chemical Physics Program and Institute for Physical Science and Technology,  University of Maryland, College Park, Maryland 20742, USA}
\author{Bei-Lok Hu\orcidlink{0000-0003-2489-9914}}
\email{blhu@umd.edu}
\affiliation{Maryland Center for Fundamental Physics and Joint Quantum Institute,  University of Maryland, College Park, Maryland 20742, USA}

\begin{abstract}

This work strives to better understand how the entanglement in an open quantum system, here represented by two coupled Brownian oscillators, is affected by a nonMarkovian  environment (with memories), here represented by two independent baths each oscillator separately interacts with.  We consider two settings, a `symmetric' configuration  wherein the parameters of both oscillators and their baths are identical,  and an `asymmetric' configuration wherein they are different,  in particular, a  `hybrid' configuration, where one of the two coupled oscillators interacts with a  nonMarkovian bath and the other with a Markovian bath. Upon finding the solutions to the Langevin equations governing the system dynamics and the evolution of the covariance matrix elements entering into its entanglement dynamics, we ask two groups of questions: 
Q1) Which time regime does the bath's nonMarkovianity benefit the system's entanglement most? The answers we get from detailed numerical studies suggest that 
A1) For an initially entangled pair of oscillators,  we see that in the intermediate time range, the duration of entanglement is proportional to the memory time, and it lasts a fraction of the relaxation time, but at late times when the dynamics reaches a steady state, the value of the symplectic eigenvalue of the partially transposed covariance matrix barely benefit from the bath nonMarkovianity. For the second group of questions: Q2) 
Can the memory of one nonMarkovian bath be passed on to another Markovian bath? And if so, does this memory transfer help to  sustain  the system's entanglement dynamics? Our results from numerical studies of the asymmetric hybrid configuration indicate that A2) A system with a short memory time can acquire improvement when it is coupled to another system with a long memory time, but, at a cost of the latter. The sustainability of the bipartite entanglement is determined by the party which breaks off entanglement most easily.  
\end{abstract}
\date{\today}

\maketitle
\hypersetup{linktoc=all}
\baselineskip=18pt
\numberwithin{equation}{section}
\allowdisplaybreaks

\newpage
\section{Introduction}

In this work we continue the inquiries two of us conducted in  2015   concerning the quantum thermodynamics and entanglement dynamics of two coupled oscillators, each with its private bath, for possible applications to  quantum processes in  systems involving two heat baths such as quantum transport and the design of quantum heat engines and other quantum devices.  In  \cite{ness} we show that for bilinear couplings between the oscillators, and between each oscillator and its own bath, described by a scalar field at temperatures $T_1>T_2$, the dynamics of the system can be solved exactly at arbitrary temperatures, even for strong coupling, thanks to the Gaussian structure. In particular, a nonequilibrium steady state (NESS) of this system can indeed exist at late times, its insensitivity to the initial conditions of the system testifies to the uniqueness of this NESS. In \cite{hotentanglemt} we studied how entanglement between the two coupled oscillators depend on the temperatures of the two baths.  We find that the valid entanglement criterion in general is not a function of the bath temperature difference, in contrast to thermal transport in the same setting. In fact lowering the temperature of one of the thermal baths does not necessarily help to safeguard the entanglement between the oscillators. Rather, quantum entanglement will disappear if any one of the thermal baths has a temperature higher than a critical temperature $T_c$.  The baths in both of these studies are made up of scalar fields which are Ohmic (linear in frequency), and, with a full spectrum, there is no imposed cutoff in its frequency range. With no particular time scale involved they are  memoryless. Scalar field baths are  Markovian.   

In this work we focus on the effects of nonMarkovian (nM) baths on the nonMarkovian dynamics of an open system in the same setup, with constant coupling between the two quantum oscillators, but not limiting our attention to very early or very late times. Rather, we aim at ranges where nonMarkovian dynamics are prominent  and  examine how the  entanglement of the system evolves. In particular, we are interested in how the memory effects associated with the nonMarkovian dynamics affects the system's entanglement throughout the evolution.  Our next paper \cite{nMhot}  will consider time-dependent coupling between the two oscillators and focus on the conditions which could allow entanglement to survive even at high temperatures, the so-called `hot entanglement' first discovered by Galve et al \cite{Galve}.  Since nonMarkovianity is the central theme of both papers we begin with a brief description of nonMarkovianity in open quantum systems. 

\subsection{nonMarkovianity in open quantum systems}

`Closed' systems in Nature do not exist, except perhaps the universe as a whole. It is a conceptual idealization. Almost all physical systems possess some environment which they interact with, no matter how weakly. Likewise, treating dynamical processes of a system as Markovian (memoryless), though commonplace, is an approximation which needs justification. In nature, rarely can any system function properly in a sustained way without memory.  

Open system approach to the description of Nature has a long history, three centuries at least, in terms of thermodynamics and statistical mechanics, where a bath characterized by a few parameters serves as the environment of a system one is interested in, and clever notions like ensembles are introduced for treating systems in equilibrium conditions or linear response theories for near-equilibrium cases.

Quantum systems add to these the factors of spin-statistics and quantum phase effects like  coherence and entanglement. The latter's  significance was first put to use in quantum optics in the 50s/60s and in quantum information sciences in the 90s/00s, both bringing forth application consequences of revolutionary magnitudes.

Placed in this light, one can see the importance of nonMarkovian quantum processes and appreciate both their theoretical and practical values, with regard to the foundations of open quantum systems and applications to the many branches of quantum information sciences and engineering.

\subsection{System's nonMarkovian dynamics linked to nonMarkovian baths}

Because of the manifold challenges unmatched by, and not encountered in, the more familiar Markovian processes, serious systematic investigations of nonMarkovian processes in open quantum systems (OQS) \cite{OQS} after three decades are still in its developmental stage. Notable landmarks were the derivation of a nonMarkovian master equation for quantum Brownian motion (QBM) \cite{HPZ} (and its associated Fokker-Planck-Wigner equation \cite{HalYu} and Langevin equation \cite{CRV}), mathematical formulations and  proposals for the criteria and measures of nonMarkovianity, and studies of nonMarkovian behavior in various schemes of quantum information processing. This subject is now also enriched by several comprehensive reviews \cite{nMrev}.

Solving the underlying nonMarkovian dynamics enables one to investigate the time-evolution of entanglement in the open system. 
For specific prior work with similar setups as here with nonMarkovianity emphasis we mention the following:   Maniscalco et al \cite{Manis}  study the non-Markovian dynamics of a two-mode bosonic system interacting with two uncorrelated thermal bosonic reservoirs and,  from it,   the dynamics of entanglement for bipartite Gaussian states. They analyze the effects of short-time system-reservoir correlations on the separability thresholds and show that the relevant parameter is the reservoir spectral density. If the frequencies of the involved modes are within the reservoir spectral density, entanglement persists for a longer time than in a Markovian channel.  Liu and Goan \cite{LiuGoan} investigate the entanglement evolution of two interacting harmonic oscillators   under the influence of non-Markovian thermal environments for both cases, namely, each oscillator has its own independent thermal reservoir, or sharing a common reservoir.  An and Zhang \cite{AnZhang}  use the rotating-wave form for the inter-oscillator coupling to investigate the entanglement dynamics of continuous-variable quantum channels in terms of an entangled squeezed state of two cavity fields in a general non-Markovian environment at zero temperature.  The influence of environments with different spectral densities, e.g., Ohmic, sub-Ohmic, and super-Ohmic, is numerically studied.  Wilson et al \cite{Wilson} study the nonMarkovian effects on the dynamics of entanglement in an exactly solvable model that involves two independent oscillators, each coupled to its own stochastic noise source.  They found that all memory effects enter via the functional form of the energy and hence the time of death and rebirth is controlled by the amount of noisy energy added into each oscillator.

We also mention in passing the other commonly studied set-up, which is to place the two coupled oscillators in a shared bath.  It is more involved due to the addition of environment-induced nonMarkovian interaction between the two oscillators. For entanglement dynamics under constant inter-oscillator coupling,  see e.g., \cite{HHPRD15,HHPLB15}, where a comparison table with  a handful oft-cited works can be found.   We plan to address hot entanglement under parametric coupling in a common bath setting in a future work. 

\subsection{System's nonMarkovian Entanglement Dynamics due to nonMarkovian Baths}

\subsubsection{nonMarkovian dynamics from nonMarkovian baths}

Instead of appealing to formal narratives or abstract proofs we would like to seek explicit solutions of the reduced system dynamics to get some direct feels of the physics. To do this we need to cover time ranges beyond the initial transient stage where most of the papers addressing the nonMarkovian criteria and measures are focused upon. Behavior of the nonMarkovian open quantum system in the mid-time-range is needed to assess how long and in what degree an initially entangled state can sustain its entanglement against environmental degradation. The initial transient stage often contains artefacts due to somewhat contrived initial conditions such as the jolt, resulting from the assumption of a product initial state.  The late time behavior is often easier to obtain from asymptotic analytical solutions, but then in most cases without any external drive, the initial entanglement is likely to have died off.

In this and a follow-up paper \cite{nMhot} we assume the same  two coupled quantum oscillators and two thermal baths set-up   as described in the two 2015 papers mentioned earlier but place our focus on how a nonMarkovian (nM) environment may alter the entanglement dynamics of the coupled oscillator system.   We will consider the simplest nonMarkovianity configuration, where the memory time of the bath is realized by the cutoff scale in the spectral density. It quantifies the time scale over which the system's evolution depends on its past history. The state of the system at earlier times is remembered by the bath, and that information is continuously fed back to the system at later moments which affects the system's subsequent evolution.

\subsubsection{nonMarkovian baths with different spectral density functions}

To identify what imparts the nonMarkovian (nM) properties of a bath note that two factors enter  in the spectral density function $J (\omega, \Lambda)$. We can go by what is described in \cite{HPZ}:

``Different environments are classified according to the functional form of the spectral density $I(\omega)$. On physical grounds, one expects the spectral density to go to zero for very high frequencies. Let us introduce a certain cutoff
frequency $\Lambda$ (a property of the environment) such that $I(\omega) \rightarrow 0$ for $\omega \gg \Lambda$. The environment is classified as
Ohmic  if in the physical range of frequencies
($\omega < \Lambda$) the spectral density is such that $I(\omega) \propto \omega$, as
supraohmic if $I(\omega) \propto \omega^n$ if $n > 1$, or as subohmic if $n < 1$. " 

Let us write the spectral density $J(\omega)$ of the bath in the frequency domain $\omega$ in  the generic form (cf,  e.g., Eq. (I.3) in \cite{HPZ})  $J(\omega)=\omega^n\mathcal{P}_{\Lambda}(\omega)$ involving two factors. First, the power of $\omega$ in it distinguishes between Ohmic and non-Ohmic cases,  as quoted above.   An environment modeled by a massless scalar field in four dimensional spacetime is an example of Ohmic  bath.     The broader class of non-Ohmic spectrum is subtler in the sense that it may induce instability in the oscillator's dynamics. An example is radiation reaction in moving charges in an electromagnetic field,  with the appearance of higher derivative terms and runaway solutions. The advantage of taking a nonMarkovian approach, treating the electromagnetic field as a supra-Ohmic bath, in mitigating these `pathologies' is discussed in a recent paper  \cite{nMRR}.  

\subsubsection{Memory in nonMarkovian baths linked to finite cut-off frequency}

The second factor, the cutoff scale  $\Lambda$, is what we focus on in this paper.  In this context,   the cutoff scale serves dual  purposes: straightforwardly it suppress the high frequency contribution of the bath modes, so we require that in the spectral density that the cut-off function $\mathcal{P}_{\Lambda}(\omega)$ should fall off to zero sufficiently fast, when $\omega$ is greater than the cutoff scale $\Lambda$. For simplicity, we assume that $\Lambda$ is the only scale in $\mathcal{P}_{\Lambda}(\omega)$. Then another requirement that $\mathcal{P}_{\Lambda}(\omega)\to 1$ as $\Lambda\to\infty$ will give the spectral density of a Markovian bath in that limit. 

When the cutoff scale in the spectral density of the bath takes on a finite value, the Langevin equation of motion, which dictates the evolution of the reduced system, will contain a  nonlocal integral expression. The nonloal expression determines how the system's evolution depends on its past history, and the cutoff scale measures the time scale this dependence lasts.  The inverse of the cutoff scale gives the memory time and this is how it is linked to the nonMarkovianity of the bath. 

Several types of cut-off functions have been analyzed in~\cite{nMRR}, namely, the Lorentzian and an exponentially decaying spectral density. The undesirable features of imposing a hard-cutoff as commonly practiced is also addressed. In this paper we suppose that $\mathcal{P}_{\Lambda}(\omega)$ has a double Lorentzian form, $\mathcal{P}_{\Lambda}(\omega)=\Lambda^4/(\Lambda^2+\omega^2)^2$. In our next paper for parametrically-driven coupled systems we shall study nonMarkovian baths of Lorentzian form, previously also investigated by Estrada and Pachon  \cite{EstPac}.

\subsubsection{Symmetric versus asymmetric set-ups}

In this work we assume that the two private baths attached to each oscillator is Ohmic and has a double-Lorentzian spectrum. 
We consider the qualitative behavior difference between two categories of parameter choices which we refer to as `symmetric' versus `asymmetric' settings or configurations.  By  a symmetric setting, we mean that both oscillators and the attached private baths have the same physical parameters, while in the asymmetric case, any of the parameters can be different. Here we focus on the asymmetric setting caused by different cutoff scales in the bath's spectral density.  For the special case of asymmetric setting when one private bath is nonMarkovian and the other is Markovian we call the reduced system  a hybrid system. 


\subsection{Issues of interest }
Here, in the two coupled oscillators with independent (uncorrelated and private) baths setting, we pose the following questions:
\begin{enumerate}
    \item How does the nonMarkovianity in the bath affect the system's entanglement ?
    \item Which time regime does bath nonMarkovianity benefit the system's entanglement most?
    \item Can the memory of one nonMarkoviann bath be passed on to another Markovian bath?
    \item Does this memory transfer help to  sustain  the system's entanglement dynamics?
\end{enumerate}
{The answers to these questions regarding the baths' nonMarkovianity and the system's entanglement dynamics can be found in the last section Sec.~\ref{S:gdiser}. Before drawing conclusions on these issues we need to proceed in three stages :  First,  provide some background and explain the characteristics of the non-Markovian dynamics for the configuration under consideration.  Then we commence our investigation with these questions in mind for the two different configurations, which we call symmetric and asymmetric. In performing numerical investigation we choose parameter ranges also with these issues in mind.  Finally,  after sufficiently useful results are obtained for the system's entanglement dynamics, we examine their behaviors, extract their meanings and attempt to  answer  the above list of questions.}

{The paper is organized as follows: We first summarize the formulation of the coupled nonMarkovian system in Sec.~\ref{S:etbksds}. Then in Sec.~\ref{S:etgisgs} we identify the characteristics of the nonMarkovian dynamics by means of numerical method. It is  most easily done in the symmetric configuration setting, where we focus on the effects of the memory time scale on the evolution of the system from solutions to the equation of motion and the evolution of the covariance matrix. This   clarifies the role of  memory effects in nonMarkovian dynamics. We then look into the entanglement dynamics from the initial moment to the stage when the coupled system has relaxed to a steady state, and pin down the time regime when the entanglement in the system benefits from the bath nonMarkovianity most. During the course of investigation, we notice some similarity in the behavior between a nonMarkovian system strongly coupled to the thermal bath and a Markovian system weakly coupled to the bath. Something similar to this has been  used by some authors as a way to `simplify' the often hard-to-grasp nonMarkovian behaviors.  In Sec.~\ref{S:esbsdfsd}, we examine and compare these two cases,  emphasizing their fundamental differences.    In Sec.~\ref{S:egdfd} we address an interesting issue in a coupled nonMarkovian system. By using an asymmetric configuration setting of a hybrid system we ask, can the memory effect be transferred between the coupled system and if so, what is the consequence to the system's entanglement? We conclude with a discussion of these issues in the last section, Sec.~\ref{S:gdiser}.}

\section{nonMarkovian dynamics}\label{S:etbksds}
Suppose we have a pair of coupled harmonic oscillators, each of which has its own private bath. Let $\chi_{i}(t)$ be the canonical position of the $i^{\text{th}}$ oscillator, whose private bath has an initial temperature $\beta_{i}^{-1}$. We further assume that two oscillators have same physical frequency $\omega_{\textsc{p}}$. The coupling strength between the oscillators is denoted by $\sigma$. In addition, the oscillators have the same mass $m$, coupling strength $e$ with each individual's bath, but each bath may not have the same cutoff scale in its spectral density.

The equations of motion in the compact matrix form is{~\cite{hotentanglemt,ness,fdr}}
\begin{equation}
	\ddot{\bm{\Xi}}(t)+\bm{\Omega}_{\textsc{b}}^{2}\cdot\bm{\Xi}(t)-\frac{e^2}{m}\int_{0}^{t}\!ds\;\bm{G}_{R,0}^{(\Phi)}(t-s)\cdot\bm{\Xi}(s)=\frac{e}{m}\,\bm{\Phi}(t)\,,\label{E:dgksbfgs}
\end{equation}
with $\bm{\Xi}(t)=\{\chi_{1}(t),\,\chi_{2}(t)\}^{T}$ and $\bm{\Phi}(t)=\{\phi_{1}(t),\,\phi_{2}(t)\}^{T}$, where $T$ denotes the matrix transpose, and
\begin{align}
	\bm{\Omega}_{\textsc{b}}^{2}&=\begin{pmatrix}\omega_{1,\textsc{b}}^{2}&\sigma\\\sigma&\omega_{2,\textsc{b}}^{2}\end{pmatrix}\,,&\bm{G}_{R,0}^{\Phi}(\tau)&=\begin{pmatrix}G_{R,0}^{(\phi_{1})}(\tau)&0\\0&G_{R,0}^{(\phi_{2})}(\tau)\end{pmatrix}=-\frac{\partial}{\partial\tau}\bm{\Gamma}^{(\Phi)}(\tau)\,.\label{E:ruthgbd}
\end{align}
On the righthand side of \eqref{E:dgksbfgs}, the operator $\bm{\Phi}$ describes a noise force due to quantum and thermal fluctuations of the bath. It satisfies the Gaussian statistics
\begin{align}
    \langle\bm{\Phi}(t)\rangle&=0\,,&\frac{1}{2}\langle\bigl\{\bm{\Phi}(t),\bm{\Phi}^T(t')\bigr\}\rangle&=\bm{G}_{H,0}^{(\Phi)}(t-t')\,.
\end{align}
The expectation value is taken with respect to the initial thermal state of the bath. The Hadamard function $\bm{G}_{H,0}^{(\Phi)}(t-t')$ is paired with the retarded Green's function of the bath $\bm{G}_{R,0}^{\Phi}(\tau)$, 
\begin{equation}
    \bm{G}_{R,0}^{\Phi}(t-t')=i\, \theta(t-t')\,\langle\bigl[\bm{\Phi}(t),\bm{\Phi}^T(t')\bigr]\rangle\,,
\end{equation}
to formulate the fluctuation-dissipation relation of the $i^{\text{th}}$ bath~\cite{fdr}, 
\begin{equation}
    \bar{\bm{G}}_{H,0}^{(\phi_i)}(\kappa)=\coth\frac{\beta_i\kappa}{2}\,\operatorname{Im}\bar{\bm{G}}_{R,0}^{(\phi_i)}(\kappa)\,,
\end{equation}
so they are also respectively called the noise kernel and the dissipation kernel. We use the convention
\begin{equation}
    f(\tau)=\int_{-\infty}^{\infty}\!\frac{d\kappa}{2\pi}\;\bar{f}(\kappa)\,e^{-i\kappa\tau}\,,
\end{equation}
to define the Fourier transformation of the function $f(\tau)$.

We introduce the kernel function $\bm{\Gamma}^{(\Phi)}(\tau)$ to isolate the contributions to renormalizations\footnote{{When we carry out integration by parts on the integral expression in \eqref{E:dgksbfgs} with the identification \eqref{E:ruthgbd}, it yields a correction to $\bm{\Omega}_{\textsc{b}}^{2}$ in \eqref{E:dgksbfgs}. When the cutoff scale is infinite as in the field theory, we conventionally call this procedure renormalization (of the frequency).}} of or the corrections to the parameters due to nonMarkovianity of the private baths. Integration by parts leads to
\begin{equation}
	\ddot{\bm{\Xi}}(t)+\bm{\Omega}_{\textsc{p}}^{2}\cdot\bm{\Xi}(t)+8\pi\gamma\int_{0}^{t}\!ds\;\bm{\Gamma}^{(\Phi)}(t-s)\cdot\dot{\bm{\Xi}}(s)+8\pi\gamma\,\bm{\Gamma}^{(\Phi)}(t)\cdot\bm{\Xi}(0)=\frac{e}{m}\,\bm{\Phi}(t)\,,\label{E:kgbdfkhdfg}
\end{equation}
with $\gamma=e^2/(8\pi m)$, and
\begin{align}
	\bm{\Omega}_{\textsc{p}}^{2}&=\begin{pmatrix}\omega_{1,\textsc{p}}^{2}&\sigma\\\sigma&\omega_{2,\textsc{p}}^{2}\end{pmatrix}\,,&\omega_{i,\textsc{p}}^{2}&=\omega_{i,\textsc{b}}^{2}-8\pi\gamma\,\Gamma^{(\phi_{i})}(0)\,.
\end{align}
Its formal solution can be formally found by the Laplace transformation of \eqref{E:kgbdfkhdfg}, and the Laplace transform of the solution is given by
\begin{equation}\label{E:bgkdgs}
	\tilde{\bm{\Xi}}(z)=\tilde{\bm{D}}_{1}(z)\cdot\bm{\Xi}(0)+\tilde{\bm{D}}_{2}(z)\cdot\dot{\bm{\Xi}}(0)+\frac{e}{m}\,\tilde{\bm{D}}_{2}(z)\cdot\tilde{\bm{\Phi}}(z)\,,
\end{equation}
with
\begin{align}\label{E:oritjsfg}
	\tilde{\bm{D}}_{2}^{-1}(z)&=z^{2}\bm{I}+\bm{\Omega}_{\textsc{p}}^{2}+8\pi\gamma\,z\,\tilde{\bm{\Gamma}}^{(\Phi)}(z)\,,&\tilde{\bm{D}}_{1}(z)&=z\,\tilde{\bm{D}}_{2}(z)\,.
\end{align}
Explicitly, $\tilde{\bm{D}}_{2}$ is given by
\begin{align}
	\tilde{\bm{D}}_{2}(z)&=\bigl[\det\tilde{\bm{D}}_{2}(z)\bigr]\begin{pmatrix}z^{2}+\omega_{2,\textsc{p}}^{2}+8\pi\gamma\,z\,\tilde{\Gamma}^{(\phi_{2})}(z)&-\sigma\\-\sigma&z^{2}+\omega_{1,\textsc{p}}^{2}+8\pi\gamma\,z\,\tilde{\Gamma}^{(\phi_{1})}(z)\end{pmatrix}\,,
\end{align}
where 
\begin{equation}
	\bigl[\det\tilde{\bm{D}}_{2}(z)\bigr]^{-1}=\det\tilde{\bm{D}}_{2}^{-1}(z)=\bigl[z^{2}+\omega_{1,\textsc{p}}^{2}+8\pi\gamma\,z\,\tilde{\Gamma}^{(\phi_{1})}(z)\bigr]\bigl[z^{2}+\omega_{2,\textsc{p}}^{2}+8\pi\gamma\,z\,\tilde{\Gamma}^{(\phi_{2})}(z)\bigr]-\sigma^{2}\,.
\end{equation}
Thus due to the nonvanishing inter-oscillator coupling $\sigma$, $\tilde{\bm{D}}_{2}$ will intermingle dynamics of both oscillators. This is better seen from the example
\begin{align}\label{E:gksdfber}
	\chi_{1}(t)&=\bigl[\bm{D}_{1}(t)\bigr]_{11}\,\chi_{1}(0)+\bigl[\bm{D}_{1}(t)\bigr]_{12}\,\chi_{2}(0)+\frac{1}{m}\,\bigl[\bm{D}_{2}(t)\bigr]_{11}\,p_{1}(0)+\frac{1}{m}\,\bigl[\bm{D}_{2}(t)\bigr]_{12}\,p_{2}(0)\notag\\
	&\qquad\qquad\qquad\qquad\qquad\qquad+\frac{e}{m}\int_{0}^{t}\!ds\;\Bigl\{\bigl[\bm{D}_{2}(t-s)\bigr]_{11}\,\phi_{1}(s)+\bigl[\bm{D}_{2}(t-s)\bigr]_{12}\,\phi_{2}(s)\Bigr\}\,.
\end{align}
Explicitly the displacement of oscillator 1 is also imparted by the thermal noise from the private bath of oscillator 2, indirectly via oscillator 2 through the inter-oscillator coupling. 

We may identify the normal modes by diagonalizing $\tilde{\bm{D}}_{2}$; however the transformation matrix to the normal modes tends to be rather complicated in the asymmetric case, and in the time domain such a transformation is not local in time, so we will not pursue this approach, except for the symmetric setting.

\section{symmetric setting}\label{S:etgisgs}
We first start with the symmetric setting. In this case, both oscillator are assumed to have the same physical frequencies $\omega_{1,\textsc{p}}=\omega_{2,\textsc{p}}=\omega_{\textsc{p}}$ and  damping constants $\gamma_1=\gamma_2=\gamma$. Likewise, both private baths will have the same power spectral densities and the cutoff scales $\Lambda_{1}=\Lambda_{2}=\Lambda$, so that their kernel functions are identical, $\Gamma^{(\phi_{1})}(\tau)=\Gamma^{(\phi_{2})}(\tau)=\Gamma^{(\phi)}(\tau)$. This setting is particularly convenient to investigate the effects of the bath's nonMarkovianity, manifested in this case by the cutoff scale or the memory time.

In the symmetric setting, the normal modes are nothing but the center-of-mass and the relative superpositions of the original two modes,
\begin{align}\label{E:kgjbskfbgs}
	\chi_{\pm}(t)=\frac{1}{\sqrt{2}}\chi_{1}(t)\pm\frac{1}{\sqrt{2}}\chi_{2}(t)\,.
\end{align}
Then we have
\begin{align}
	\ddot{\chi}_{\pm}(t)+\omega_{\pm}^{2}\,\chi_{\pm}(t)+8\pi\gamma\int_{0}^{t}\!ds\;\Gamma^{(\phi)}(t-s)\,\dot{\chi}_{\pm}(s)+8\pi\gamma\,\Gamma^{(\phi)}(t)\,\chi_{\pm}(0)&=\frac{e}{m}\,\phi_{\pm}(t)\,,\label{E:eotugbsd1}
\end{align}	
where $\omega_{\pm}^{2}=\omega^2\pm\sigma$ represents the oscillating frequencies of the normal modes, and $\phi_{\pm}$ is given by
\begin{equation}
    \phi_{\pm}(t)=\frac{1}{\sqrt{2}}\phi_{1}(t)\pm\frac{1}{\sqrt{2}}\phi_{2}(t)\,.
\end{equation}    
Here we remind that both private baths, associated with each oscillator, are initially uncorrelated and do not have direct coupling. Thus each mode acts as an Ohmic, nonMarkovian oscillator with different oscillating frequencies. Since for the equation of motion of the form \eqref{E:eotugbsd1} takes the generic form
\begin{equation}\label{E:eituge}
	\ddot{\chi}(t)+\omega^{2}\,\chi(t)+8\pi\gamma\int_{0}^{t}\!ds\;\Gamma^{(\phi)}(t-s)\,\dot{\chi}(s)+8\pi\gamma\,\Gamma^{(\phi)}(t)\,\chi(0)=\frac{e}{m}\,\phi(t)\,,
\end{equation}
we will first use this equation to discuss the effects of the bath nonMarkovianity on the system's dynamics.

The general solution of $\chi(t)$ is given by the inverse Laplace transform of
\begin{equation}\label{E:trbgwi}
	\tilde{\chi}(z)=\tilde{d}_{1}(z)\,\chi(0)+\tilde{d}_{2}(z)\,\dot{\chi}(0)+\frac{e}{m}\,\tilde{d}_{2}(z)\,\tilde{\phi}(z)\,,
\end{equation}
with $\tilde{d}_{1}(z)=z\,\tilde{d}_{2}(z)$, and
\begin{align}
	\tilde{d}_{2}(z)&=\frac{1}{z^{2}+\omega^{2}+8\pi\gamma\,z\,\tilde{\Gamma}^{(\phi)}(z)}\,,
\end{align}
in which
\begin{equation}
	\tilde{\Gamma}^{(\phi)}(z)=\begin{cases}
			\dfrac{1}{4\pi}\dfrac{\Lambda}{\Lambda+z}\,,&\text{Lorentzian spectral density}\,,\\[10pt]
			\dfrac{1}{8\pi}\dfrac{\Lambda(2\Lambda+z)}{(\Lambda+z)^{2}}\,,&\text{Double Lorentzian spectral density}\,.
	\end{cases}
\end{equation}
We will choose the double Lorentzian form for the bath spectral density. The two fundamental solutions $d_1(t)$ and $d_2(t)$ are two particularly useful homogeneous solutions to the equation of motion \eqref{E:eituge}. They satisfy the initial conditions $d_1(0)=1$, $\dot{d}_1(0)=0$, $d_2(0)=0$ and $\dot{d}_2(0)=1$, and are used to construct the complete solution to \eqref{E:eituge}.

\begin{figure}
\centering
    \scalebox{0.3}{\includegraphics{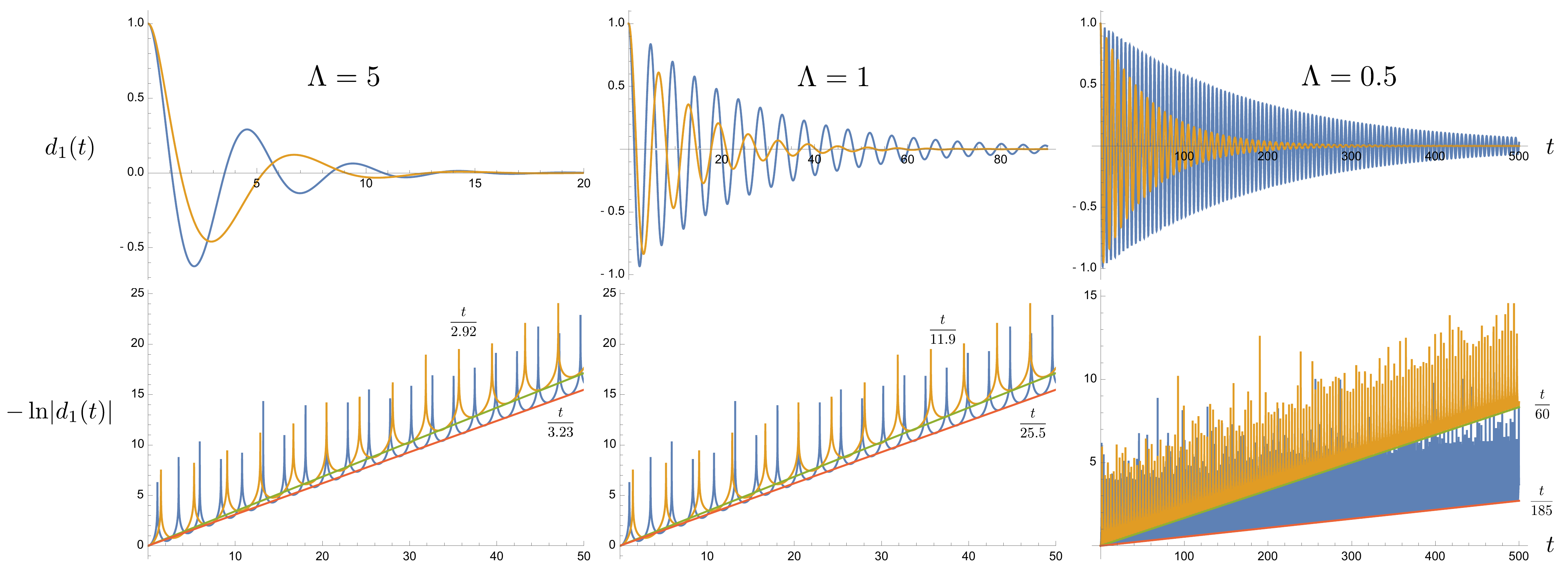}}
    \caption{The effective damping of the oscillator, coupled to a thermal bath with the double Lorentzian spectral density. The first row show the temporal behavior of $d_{1}(t)$, while the second row is that of $-\ln\lvert d_{1}(t)\rvert$. The blue curve represents the $\omega_{\textsc{p}}=1.2$ case, and the orange curve the $\omega_{\textsc{p}}=0.8$ case. The effective damping constant is smaller for a larger $\omega_{\textsc{p}}$. The parameter $\gamma$ is chosen to be 0.3.}\label{Fi:effDamp}
\end{figure}

For the double Lorentzian bath spectral density, the Laplace transform $\tilde{d}_{2}(z)$ has four poles: two negative real poles and two complex poles, whose real parts are real. Since for a given set of $\gamma$, $\omega_{\textsc{p}}$, and $\Lambda$, the two real poles are more negative than the real parts of the complex poles, so at late times, the damping behavior of $d_{2}(t)$ is controlled by the real part of the complex poles. Thus, we may construct  an effective damping constant, which is approximately given by 
\begin{equation}\label{E:fgskd}
	\gamma_{\textsc{eff}}\sim\gamma\biggl[\frac{\Lambda^{4}}{(\Lambda^{2}+\omega_{\textsc{p}}^{2})^{2}}+\frac{\gamma\Lambda^{5}(3\Lambda^{4}-7\Lambda^{2}\omega_{\textsc{p}}^{2}-2\omega_{\textsc{p}}^{4})}{(\Lambda^{2}+\omega_{\textsc{p}}^{2})^{5}}+\cdots\biggr]\,,
\end{equation}
when $\gamma/\Lambda\lesssim1$, $\gamma/\omega_{\textsc{p}}\lesssim1$ in the long memory time limit. It implies that these two modes will have two different effective damping constants due to the difference in the normal mode frequencies, even though they have the same bath configuration.

Fig.~\ref{Fi:effDamp} shows how the effective damping constant depends the oscillating frequency and the cutoff scale. We can immediately identify two features: 1) for a given cutoff scale, the damping is weaker, that is, smaller effective damping constant, when the frequency is larger. It is clearly seen in the second row of Fig.~\ref{Fi:effDamp} , and the difference is more significant for smaller cutoff scales. This feature is related to the second feature: 2) for a given oscillating frequency, the effective damping constant $\gamma_{\textsc{eff}}$  decreases if we lower the cutoff scale $\Lambda$, and  such change is more dramatic when $\Lambda$ is smaller than the oscillating frequency $\omega_{\textsc{p}}$. The weakening of the effective damping constant can be understood from two aspects. Since the inverse cutoff scale can be interpreted as the memory time of the bath, which moderates the duration the system, with which the bath interacts, depends on its past history. From this viewpoint, when the memory time is longer than the typical period $2\pi/\omega_{\textsc{p}}$ of the oscillatory motion, or when the cutoff scale $\Lambda$ is smaller than the oscillating frequency $\omega_{\textsc{p}}$, the system dynamics from the previous cycle is still in good coherence with the current one. Thus the motion is progressively superposed to compete against decaying due to dissipation. This effect is expected to be more prominent when the memory is much longer, because the contributions from more earlier cycles will coherently join in~\cite{nMRR}. Alternatively, the weakening effect can also be understood from resonance absorption~\cite{horhammer}. The equation of motion \eqref{E:eituge} more or less describes a driven, damped harmonic oscillator, although this may not be obvious when the nonlocal integral expression is present. However, we may start from the limit $\Lambda\to\infty$, where the integral expression gives $2\gamma\dot{\chi}(t)$ and Eq.~\eqref{E:eituge} reduces to the standard form. It is known that such a system has peaked power spectrum, centered at around $\omega\sim\omega_{\textsc{p}}$ with a width of the order $\mathcal{O}(\gamma)$. Now heuristically suppose we only allow bath modes whose frequencies are lower than the location of the resonance peak, $\omega_{\textsc{p}}$. These modes will not be very efficient in exchanging energy between the oscillator and the bath, so neither can they effectively drive the oscillator, nor can drain the energy out of the oscillator. Therefore it leads to relatively weak damping to the oscillator's motion.

By these arguments, we expect that for a fixed $\Lambda$, in particular when $\Lambda<\omega_{\textsc{p}}$, a larger value of $\omega_{\textsc{p}}$ means that the bath modes which participate in the interactions are further away from the resonance peak, thus rendering less effective energy exchange. Or, when the memory time becomes even longer than the system's period, thus allowing more coherent cycles from the past history of the oscillator's motion. These make the effective damping weaker.

\begin{figure}
\centering
    \scalebox{0.6}{\includegraphics{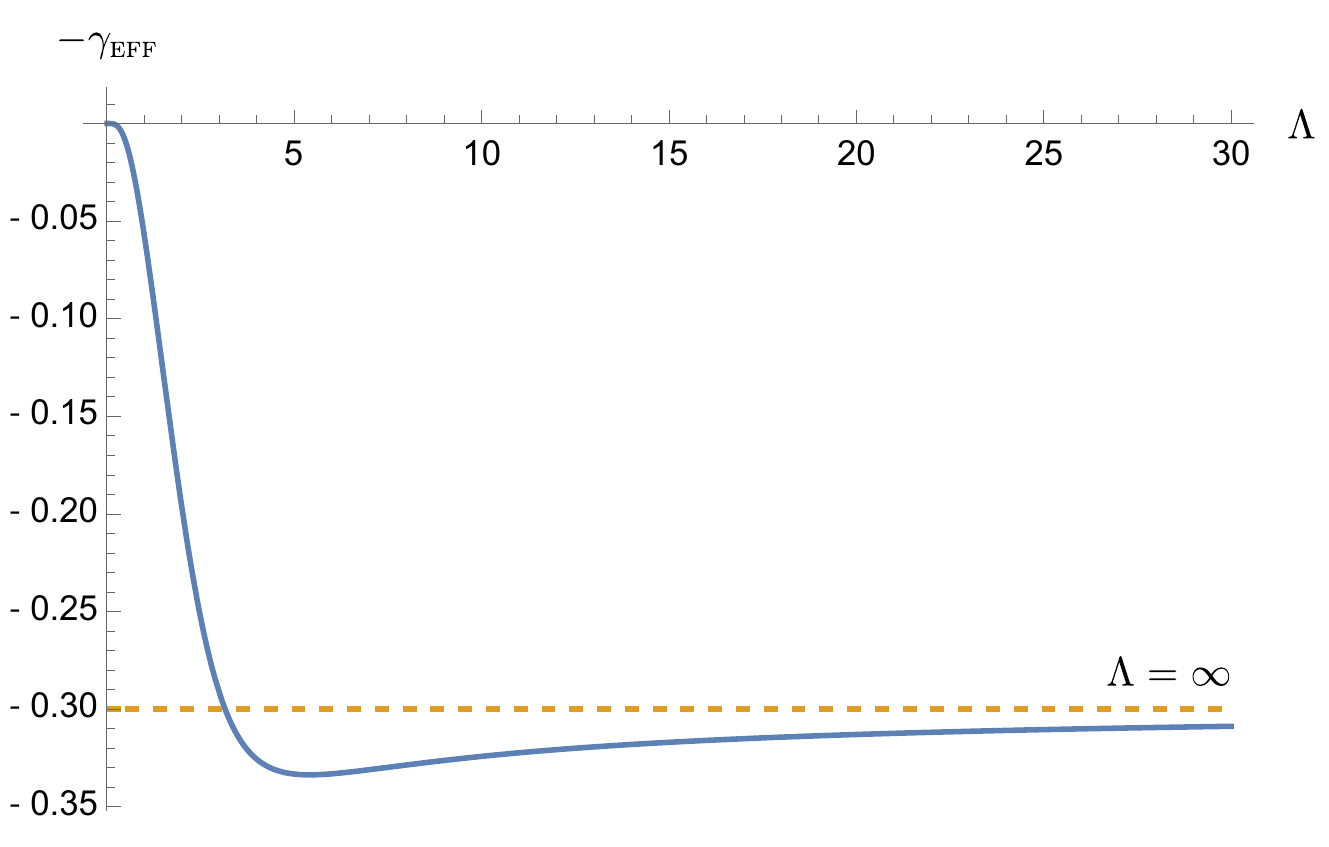}}
    \caption{The $\Lambda$ dependence of the effective damping constant $\gamma_{\textsc{eff}}$ of the oscillator, coupled to a thermal bath with the double Lorentzian spectral density. Here we choose $\omega_{\textsc{p}}=1\,\omega_{\textsc{p}}$, and $\gamma=0.3$. In this case when $\Lambda<3.57\,\omega_{\textsc{p}}$, the effective damping constant is already weaker than $\gamma$.}\label{Fi:effDampConst}
\end{figure}

Fig.~\ref{Fi:effDampConst} shows the cutoff scale dependence of the effective damping constant by numerical method. We see that when $\Lambda\lesssim 3\,\omega_{\textsc{p}}$, the effective damping constant $\gamma_{\textsc{eff}}$ monotonically decreases with $\Lambda$, and is smaller than its Markovian counterpart $\gamma=0.3\,\omega_{\textsc{p}}$. On the other hand, in this case when $\Lambda$ is greater than $20\,\omega_{\textsc{p}}$, the nonMarkovian effect is barely present. This raises an interesting observation. Here we choose $\gamma=0.3\,\omega_{\textsc{p}}$, which typically falls in the strong oscillator-bath coupling regime. By introducing nonMarkovianity via engineering the bath's spectrum, we seem to effectively render the reduced dynamics of the oscillator to behave like a weak coupling case. Or, by tuning the bath spectrum, we may be able to vary the effective coupling between the oscillator and the bath.

Based on the above arguments and the idea of effective damping constant, we may be led to ask an important strategic question: Can we find a weak-coupling Markovian effective system to mimic the strong-coupling nonMarkovian system? It can greatly facilitate the theoretical study of the nonMarkovian system, because the latter is computationally challenging and demanding. In contrast, the former is analytically and exactly solvable. The answer depends. We will come back to this in Sec.~\ref{S:esbsdfsd}.

To compute the entanglement measure\footnote{Relevant material can be found in App.~\ref{S:jgvsf}} via negativity, we start from the covariance matrix elements of the normal modes, since each mode is ``decoupled''~\footnote{Strictly speaking this is not entirely correct because $\phi_{\pm}$ are not independent.}. For the discussion of sustainability of entanglement, we suppose two oscillators are initially prepared in a two-mode squeezed vacuum state (see App.~\ref{S:kgbdf} for the discussion of the two-mode squeezed state). When the squeeze parameter $\eta$ is not zero, the initial state is already entangled, and the degree of entanglement grows with increasing $\eta$. With respect to the normal modes, the nonzero covariance matrix elements of the coupled oscillator in this initial state are
\begin{align}
    \sigma_{\chi_{+}\chi_{+}}(0)&=\dfrac{1}{2m\omega_{\textsc{p}}}e^{-2\eta}\,,&\sigma_{p_{+}p_{+}}(0)&=\dfrac{m\omega_{\textsc{p}}}{2}e^{+2\eta}\,,&\sigma_{\chi_{-}\chi_{-}}(0)&=\dfrac{1}{2m\omega_{\textsc{p}}}e^{+2\eta}\,,&\sigma_{p_{-}p_{-}}(0)&=\dfrac{m\omega_{\textsc{p}}}{2}e^{-2\eta}\,.
\end{align}
These two conditions ensure the covariance matrix with respect to the normal mode remains blockwise diagonal at all times
\begin{equation}\label{E:ktrvsj}
	\bm{\sigma}_{\pm}(t)=\begin{pmatrix}\sigma_{\chi_{+}\chi_{+}}(t) &\sigma_{\chi_{+}p_{+}}(t) &0 &0 \\[4pt]\sigma_{\chi_{+}p_{+}}(t) &\sigma_{p_{+}p_{+}}(t) &0 &0\\[4pt]0 &0 &\sigma_{\chi_{-}\chi_{-}}(t) &\sigma_{\chi_{-}p_{-}}(t)\\[4pt] 0 &0 &\sigma_{\chi_{-}p_{-}}(t) &\sigma_{p_{-}p_{-}}(t) \end{pmatrix}\,.
\end{equation}
Thus we only need to compute a smaller set of covariance matrix elements,
\begin{align}
	\sigma_{\chi_{\pm}\chi_{\pm}}(t)&=d_{1}^{(\pm)2}(t)\,\sigma_{\chi_{\pm}\chi_{\pm}}(0)+\frac{1}{m^{2}}\,d_{2}^{(\pm)2}(t)\,\sigma_{p_{\pm}p_{\pm}}(0)\notag\\
	&\qquad\qquad\qquad\qquad+\frac{e^{2}}{m^{2}}\int_{0}^{t}\!ds\!\int_{0}^{t}\!ds'\;d_{2}^{(\pm)}(t-s)\,d_{2}^{(\pm)}(t-s')\,G_{H,0}^{(\pm\pm)}(s,s')\,,\label{E:dfgks1}\\
	\sigma_{\chi_{\pm}p_{\pm}}(t)&=m\,d_{1}^{(\pm)}(t)\,\dot{d}_{1}^{(\pm)}(0)\,\sigma_{\chi_{\pm}\chi_{\pm}}(t)+\frac{1}{m}\,d_{2}^{(\pm)}(t)\,\dot{d}_{2}^{(\pm)2}(t)\,\sigma_{p_{\pm}p_{\pm}}(0)\notag\\
	&\qquad\qquad\qquad\qquad+\frac{e^{2}}{m}\int_{0}^{t}\!ds\!\int_{0}^{t}\!ds'\;d_{2}^{(\pm)}(t-s)\,\dot{d}_{2}^{(\pm)}(t-s')\,G_{H,0}^{(\pm\pm)}(s,s')\,,\\
	\sigma_{p_{\pm}p_{\pm}}(t)&=m^{2}\dot{d}_{1}^{(\pm)2}(t)\,\sigma_{\chi_{\pm}\chi_{\pm}}(0)+\dot{d}_{2}^{(\pm)2}(t)\,\sigma_{p_{\pm}p_{\pm}}(0)\notag\\
	&\qquad\qquad\qquad\qquad+e^{2}\int_{0}^{t}\!ds\!\int_{0}^{t}\!ds'\;\dot{d}_{2}^{(\pm)}(t-s)\,\dot{d}_{2}^{(\pm)}(t-s')\,G_{H,0}^{(\pm\pm)}(s,s')\,.
\end{align}
Here $d_i^{(\pm)}$ are the counterparts of the fundamental solutions, discussed in \eqref{E:trbgwi}, of the normal modes.

Then we can restore the covariance matrix elements of the canonical variables of two modes by suitable superpositions of the covariance matrix elements of the normal modes. This step is essential because entanglement is partition dependent~\cite{qic}. The entanglement measure, negativity, of a bipartite system depends on how we partition the bipartite system. Its values vary if we use the normal modes, instead of the canonical variables of the bipartite system, to construct the measure. This can be traced to the fact that the partial transposition does not belong to the symplectic transformation (see App.~\ref{S:jgvsf} for a brief discussion.)

\begin{figure}
\centering
    \scalebox{0.4}{\includegraphics{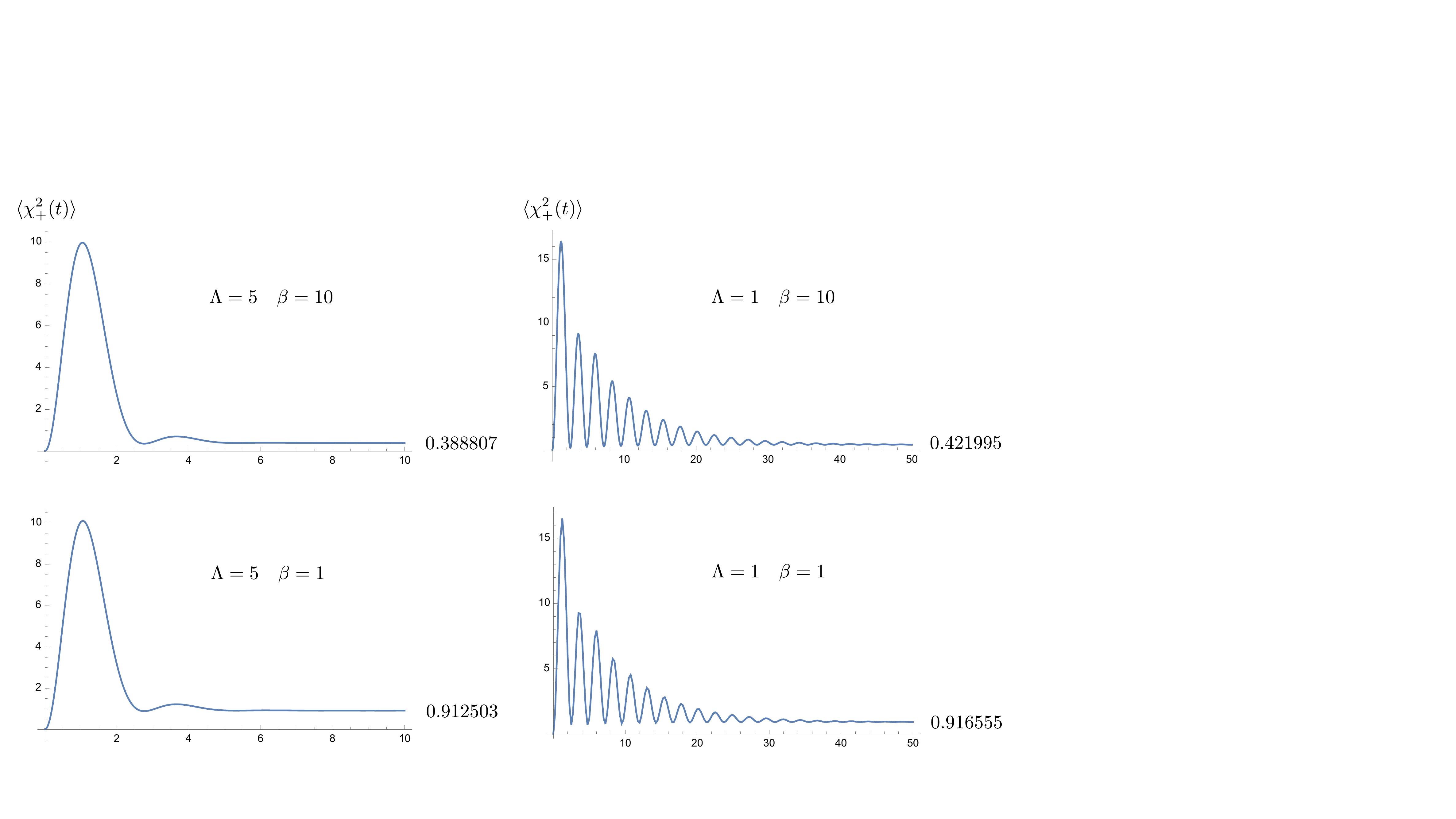}}
    \caption{The time evolution of $\langle\chi_{+}^{2}(t)\rangle$. The columns show its dependence on the bath temperature for different choices of cutoff scales, while the rows reveal the cutoff dependence. Here we choose the parameters in the unit of $\omega_{\textsc{p}}$, and they take on the values $m=1$, $\gamma=0.5$, and $\sigma=0.2$.}\label{Fi:coVmemoryTemp}
\end{figure}
Before we proceed, we take a look at the time evolution of the selected covariance matrix element $\langle\chi_{+}^{2}(t)\rangle$ for different choices of the bath temperatures and the bath cutoff scales. In Fig~\ref{Fi:coVmemoryTemp}, different rows correspond to different initial bath temperatures $\beta^{-1}$, while different columns are associated with different bath memory times $\Lambda^{-1}$. We immediately see that $\langle\chi_{+}^{2}(t)\rangle$ relaxes more slowly for longer memory times, consistent with the time evolution of the solution shown in Fig.~\ref{Fi:effDamp}. 

We observe that each covariance matrix element can be decomposed into two components: 1) the intrinsic {(active)} component depends on the initial condition of the oscillator, and is damped by {dissipation to} the bath, and 2) the induced {(passive)} component is generated by the {quantum and thermal fluctuations from} the bath, independent of the oscillator's initial condition. The latter is the only component that will survive at late times, so the late-time value of the covariance matrix elements will not depend on the initial conditions of the system. Both components obey different statistics and are not correlated for the linear system. The memory endowed by the bath has different effects in both components. In the intrinsic component, the nonMarkovian effect leads to a weaker damping rate, so we expect that the information on the initial conditions of the system will linger for an extended period of time. On the other hand, in the passive component, the coherent superposition of the oscillator's motion over previous cycles, due to the memory effect, will compete with the accumulative intervention, resulting from the thermal fluctuations of the bath. 

In each column, when the intrinsic component dies down, the value of the $\langle\chi_{+}^{2}(t)\rangle$ dispersion will not decay to zero; otherwise it will violate uncertainty principle. In fact it will more of less come a constant due to balance between the thermal fluctuations and damping, and this constant value grows with the bath temperature because at higher temperatures, the bath fluctuations are stronger and becomes more random, less correlated, but damping in this case is independent of the bath temperature.  

\begin{figure}
\centering
    \scalebox{0.45}{\includegraphics{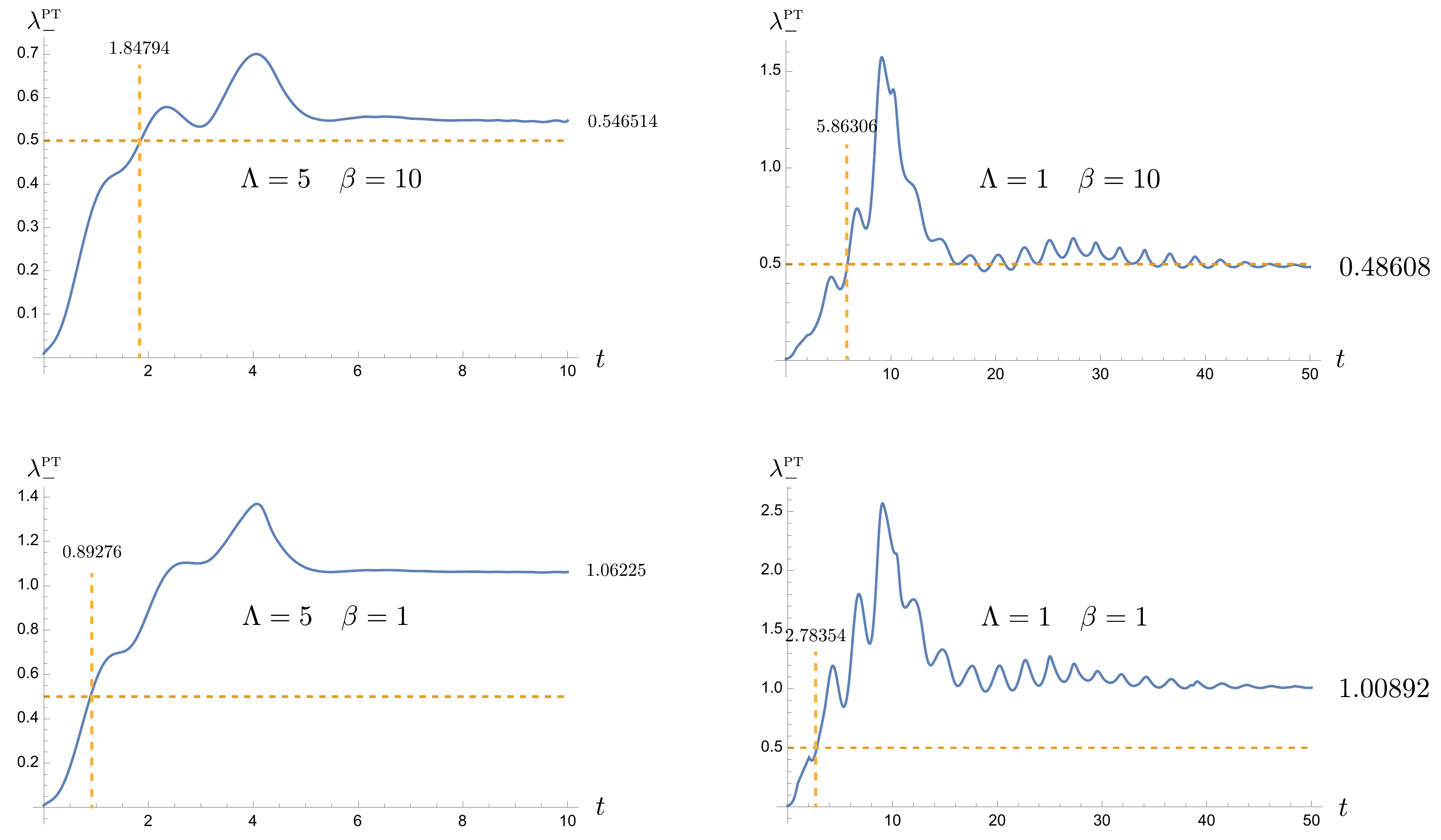}}
    \caption{The time evolution of the symplectic eigenvalue $\lambda^{\textsc{pt}}_{-}(t)$. The advantage of the nonMarkovian, memory effect is clearly seen during the intermediate time regime. The entanglement is present when $\lambda_-^{\textsc{pt}}(t)<1/2$. The parameters are chosen to have $m=1$, $\gamma=0.5$, and $\sigma=0.2$ in the unit of $\omega_{\textsc{p}}$.}\label{Fi:enTmemoryTemp}
\end{figure}
A longer memory time implies a longer effective relaxation time scale, and thus a longer duration for the effects of thermal fluctuations to accumulate. This probably explains the larger values of the dispersions $\langle\chi_+^2(t)\rangle$ at late times when the cutoff scale $\Lambda$ is smaller in each row of Fig.~\ref{Fi:coVmemoryTemp}. However, this difference is rather insignificant at higher bath temperature, signifying the ineffectiveness of nonMarkovian effects at higher temperatures.

We will transform the covariance matrix $\bm{\sigma}_{\pm}(t)$ back to that with respect to the modes $\chi_{1,2}$ by
\begin{align}
	\bm{\sigma}(t)&=\bm{U}_{2}^{\vphantom{T}}\cdot\bm{\sigma}_{\pm}(t)\cdot\bm{U}_{2}^{T}\,., &\bm{U}_{2}^{-1}=\bm{U}_{2}&=\begin{pmatrix}\frac{1}{\sqrt{2}}&0&\frac{1}{\sqrt{2}}&0\\[4pt]0 &\frac{1}{\sqrt{2}} &0 &\frac{1}{\sqrt{2}}\\[4pt]\frac{1}{\sqrt{2}} &0 &-\frac{1}{\sqrt{2}} &0\\[4pt]0&\frac{1}{\sqrt{2}}&0&-\frac{1}{\sqrt{2}}\end{pmatrix}\,,
\end{align}
where $\bm{U}_{2}$ is a global symplectic matrix, that is, $\bm{U}_{2}\in\operatorname{Sp}(4,\mathbb{R})$.

The symplectic eigenvalues of the partially transposed covariance matrix $\bm{\sigma}^{\textsc{pt}}$ is then given by
\begin{align}\label{E:gbdks}
	(\lambda^{\textsc{pt}}_{\pm})^{2}&=\frac{1}{2}\Delta(\bm{\sigma}^{\textsc{pt}})\pm\frac{1}{2}\sqrt{\Delta^{2}(\bm{\sigma}^{\textsc{pt}})-4\det\bm{\sigma}^{\textsc{pt}}}\,.\,
\end{align}
with $\Delta(\bm{\sigma}^{\textsc{pt}})$, $\det\bm{\sigma}^{\textsc{pt}}$ spelled out below in terms of covariance matrix elements of the normal modes as
\begin{align}
        \Delta(\bm{\sigma}^{\textsc{pt}})&=\sigma_{\chi_{+}\chi_{+}}(t)\times\sigma_{p_{-}p_{-}}(t)+\sigma_{\chi_{-}\chi_{-}}(t)\times\sigma_{p_{+}p_{+}}(t)-2\sigma_{\chi_{+}p_{+}}(t)\sigma_{\chi_{-}p_{-}}(t)\,,\\
	\det\bm{\sigma}^{\textsc{pt}}=\det\bm{\sigma}&=\bigl[\sigma_{\chi_{+}\chi_{+}}(t)\times\sigma_{p_{+}p_{+}}(t)-\sigma_{\chi_{+}p_{+}}^{2}(t)\bigr]\bigl[\sigma_{\chi_{-}\chi_{-}}(t)\times\sigma_{p_{-}p_{-}}(t)-\sigma_{\chi_{-}p_{-}}^{2}(t)\bigr]\,.
\end{align}
Thus, the entanglement of the system at any given moment $t$ is determined by the value $\lambda^{\textsc{pt}}_{-}(t)<1/2$. 

The time evolution of the symplectic eigenvalue $\lambda^{\textsc{pt}}_{-}$ is shown in Fig.~\ref{Fi:enTmemoryTemp}. Since it contains products of various elements of the covariance matrix, we cannot simply decompose $\lambda^{\textsc{pt}}_{-}$ into the intrinsic and the induced components as we did for each covariance matrix elements. However, we still expect that the initial entanglement between the oscillators will be gradually lost because contributions related to the oscillators' initial conditions will be exponentially small, and that the late-time behavior of the symplectic eigenvalue $\lambda^{\textsc{pt}}_{-}$ is predominantly controlled by the bath. Thus the existence of the entanglement at late times will be independent of the initial conditions of the oscillators, but determined by the configuration of the bath such as the memory time, temperature and the oscillator-bath coupling strength.

\begin{figure}
\centering
    \scalebox{0.5}{\includegraphics{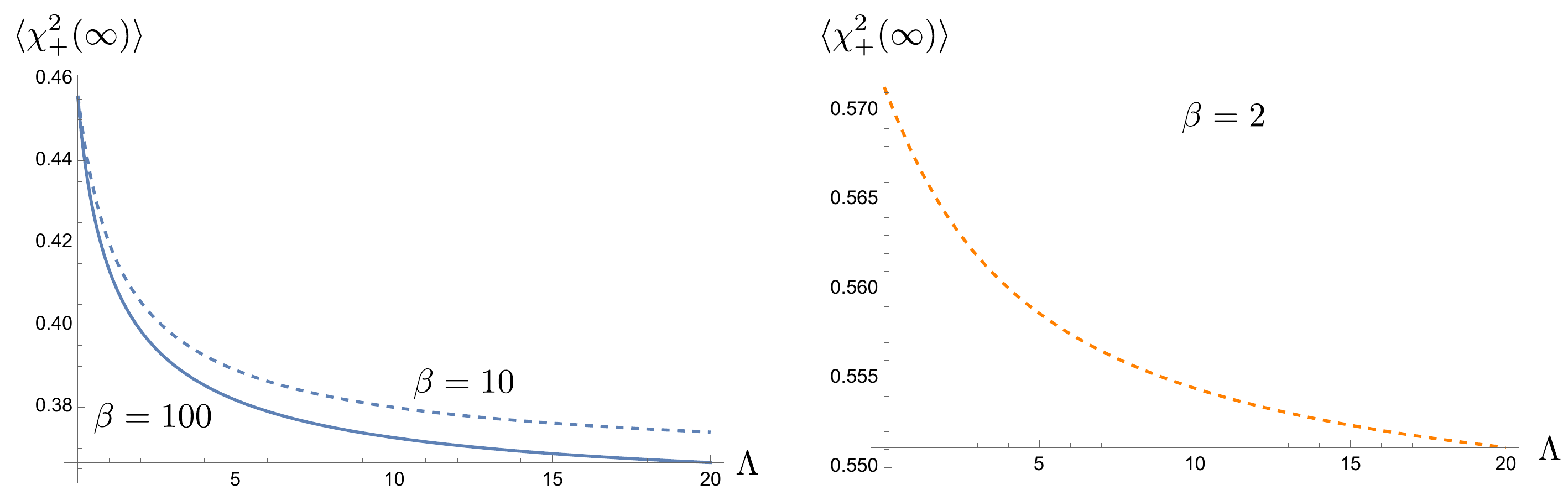}}
    \caption{The cutoff dependence of the element $\sigma_{\chi_{+}\chi_{+}}(\infty)$. The memory effect is suppressed at high bath temperatures. We choose $\omega_{\textsc{p}}=1$, $m=1$, $\gamma=0.5$, and $\sigma=0.2$.}\label{Fi:entLTvsLamb}
\end{figure}

Since we choose an entangled initial state, we see that the memory effect is much more significant during the relaxation stage. The duration $\tau_{\textsc{ent}}$ for both oscillators to remain entangled is correlated with the length of memory time. In Fig.~\ref{Fi:enTmemoryTemp} the coupled oscillators become separable when $\lambda_-^{\textsc{pt}}\geq1/2$. Comparing the plots in each row, we observe that the entanglement interval during the relaxation stage increases with the memory time $\tau_{\textsc{mem}}$. In other words, the entanglement is more robust against the thermal fluctuations and can be sustained at higher bath temperatures, owing to {the nonMarkovian effect due to the bath's memory}. This point is particularly clearly seen when we compare the $\Lambda=5\,\omega_{\textsc{p}}$, $\beta=10\,\omega_{\textsc{p}}^{-1}$ and $\Lambda=1\,\omega_{\textsc{p}}$, $\beta=1\,\omega_{\textsc{p}}^{-1}$ cases. In the latter, the bath temperature has 10 times higher than the former, but the entanglement dies out at approximately the same time around $t=2\,\omega_{\textsc{p}}^{-1}$, as shown in Fig.~\ref{Fi:enTmemoryTemp}. It is expected to be more dramatic for even smaller cutoff scale, as is inferred from the behavior of oscillator dynamics, shown in Fig.~\ref{Fi:effDamp}. The result that $\tau_{\textsc{ent}}\sim\mathcal{O}(1)\,\omega_{\textsc{p}}^{-1}$ may not seem significant as it appears. However, recall that the oscillators are  strongly coupled to the bath because $\gamma=0.5\,\omega_{\textsc{p}}$. In Fig.~\ref{Fi:asymSymp}, we will see that $\tau_{\textsc{ent}}$ for the corresponding Markovian bath (green curve) is merely $\tau_{\textsc{ent}}\simeq 0.197\,\omega_{\textsc{p}}^{-1}$. Furthermore, let us put it in a comparative perspective. For example, $\tau_{\textsc{ent}}\simeq 5.86\,\omega_{\textsc{p}}^{-1}$ for the $\Lambda=1\,\omega_{\textsc{p}}$, $\beta=1\,\omega_{\textsc{p}}$ case. The effective damping constants for the two normal modes are approximately given by $\gamma^{(+)}_{\textsc{eff}}\simeq0.066\,\omega_{\textsc{p}}$ and $\gamma^{(-)}_{\textsc{eff}}\simeq0.094\,\omega_{\textsc{p}}$. Thus the effective relaxation time is roughly $\tau_{\textsc{r}}\simeq10\sim15\,\omega_{\textsc{p}}^{-1}$. The memory time in this case is $\tau_{\textsc{mem}}=1\,\omega_{\textsc{p}}$. Thus the entanglement duration is about six-fold of the memory time and one third of relaxation time. Thus, this entanglement duration is quite appreciable in the relaxation process. It can be further improved if we enhance the initial squeezing.

In Fig.~\ref{Fi:enTmemoryTemp}, we also see that although the oscillators become disentangled after $\tau_{\textsc{ent}}$, determined by $\omega_{\textsc{p}}$, $\gamma$, $\Lambda$ and $\beta$, the entanglement can resurrect at a later time, when the memory time is sufficiently long and the bath temperature is low enough, as shown in the case $\Lambda=1$, $\beta=10$. However, this late-time entanglement does not benefit from the initial squeezing because its effect has been dissipated away. The plots show that although the nonMarkovian effect may improve entanglement at late times, the benefit is marginal. To understand this better, let us focus on the behavior of the covariance matrix elements and the symplectic eigenvalue $\eta_{-}^{\textsc{pt}}$ after the oscillators are relaxed to the {steady} state.

\begin{figure}
\centering
    \scalebox{0.35}{\includegraphics{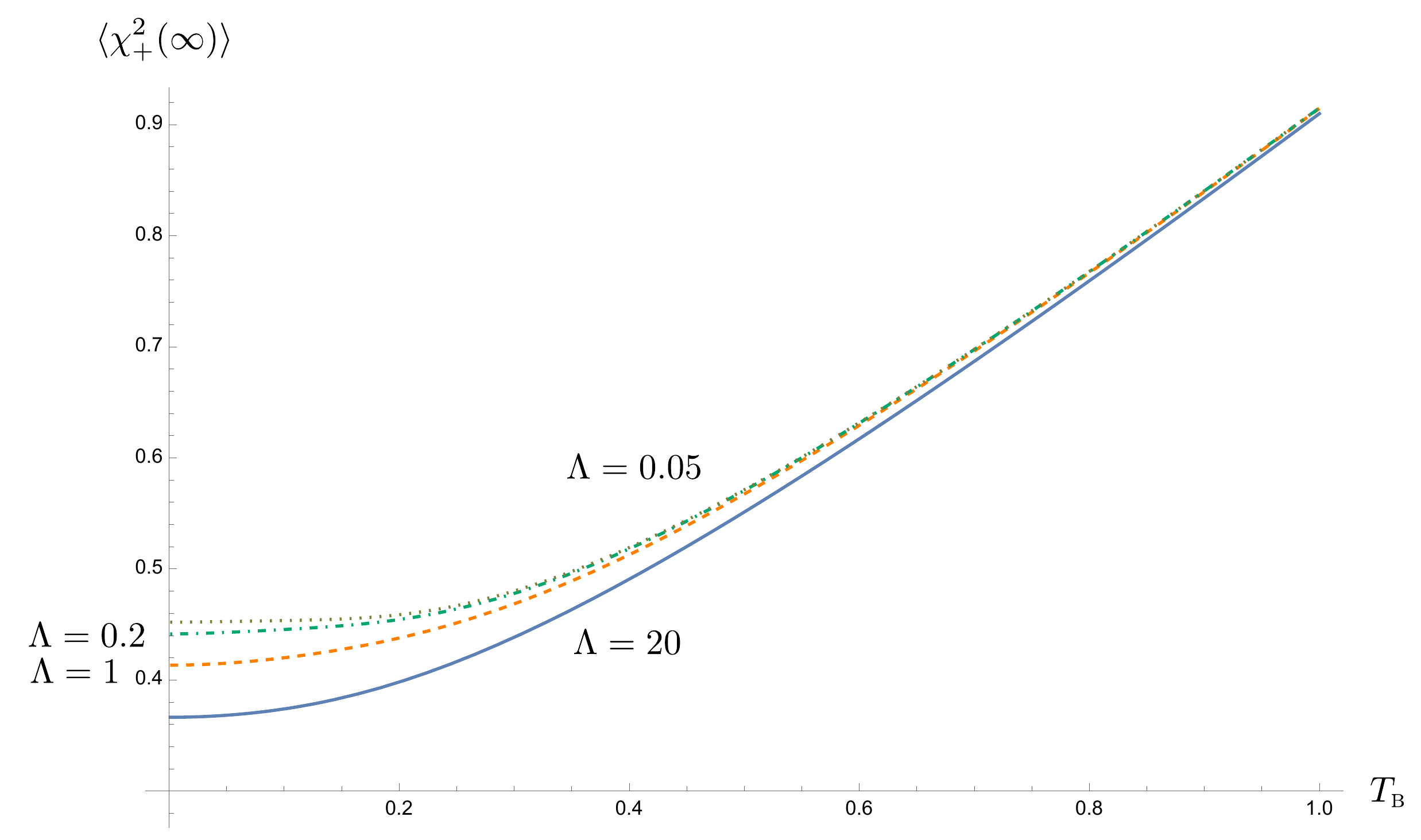}}
    \caption{The temperature dependence of the late-time value of $\sigma_{\chi_{+}\chi_{+}}(t)$. For a given bath temperature, $\sigma_{\chi_{+}\chi_{+}}(\infty)$ tends to be larger for the longer memory time. Here the parameters take on the values $m=1$, $\gamma=0.5$, and $\sigma=0.2$ in the unit of $\omega_{\textsc{p}}$}\label{Fi:entLTvsTemp}
\end{figure}

The late-time result in this case is particularly simple, since $\sigma_{\chi_{+}p_{+}}(\infty)$ and  $\sigma_{\chi_{-}p_{-}}(\infty)$ vanish. We are left with only four elements $\sigma_{\chi_{\pm}\chi_{\pm}}(\infty)$ and $\sigma_{p_{\pm}p_{\pm}}(\infty)$, and they are given by
\begin{align}
	\sigma_{\chi_{\pm}\chi_{\pm}}(\infty)&=\frac{1}{m}\int_{-\infty}^{\infty}\!\frac{d\kappa}{2\pi}\;\coth\frac{\beta\kappa}{2}\,\operatorname{Im}\bar{d}_{2}^{(\pm)}(\kappa)\,,\\
	\sigma_{p_{\pm}p_{\pm}}(\infty)&=m\int_{-\infty}^{\infty}\!\frac{d\kappa}{2\pi}\;\kappa^{2}\coth\frac{\beta\kappa}{2}\,\operatorname{Im}\bar{d}_{2}^{(\pm)}(\kappa)\,.
\end{align}
Fig.~\ref{Fi:entLTvsLamb} shows that the late-time value of $\sigma_{\chi_{+}\chi_{+}}(t)$ decreases with the larger cutoff scale $\Lambda$ or the shorter memory time. It is consistent with earlier observations, and it can be clearly seen that the variation of $\sigma_{\chi_{+}\chi_{+}}(t)$ is less significant, implying that the nonMarkovian effect is much weaker at higher bath temperature. This is also demonstrated in Fig.~\ref{Fi:entLTvsTemp}, where the curves corresponding to different cutoff scales essentially converge when the bath temperature $T_{\textsc{b}}$ is of the order $\omega_{\textsc{p}}$. Note that the late-time values of the covariance matrix elements are mainly governed by the bath. When the bath has a higher initial temperature $\beta^{-1}$, it imparts stronger thermal fluctuations to the system. Thus the values of $\sigma_{\chi_{+}\chi_{+}}(\infty)$ quickly increase with the bath temperature.

\begin{figure}
\centering
    \scalebox{0.4}{\includegraphics{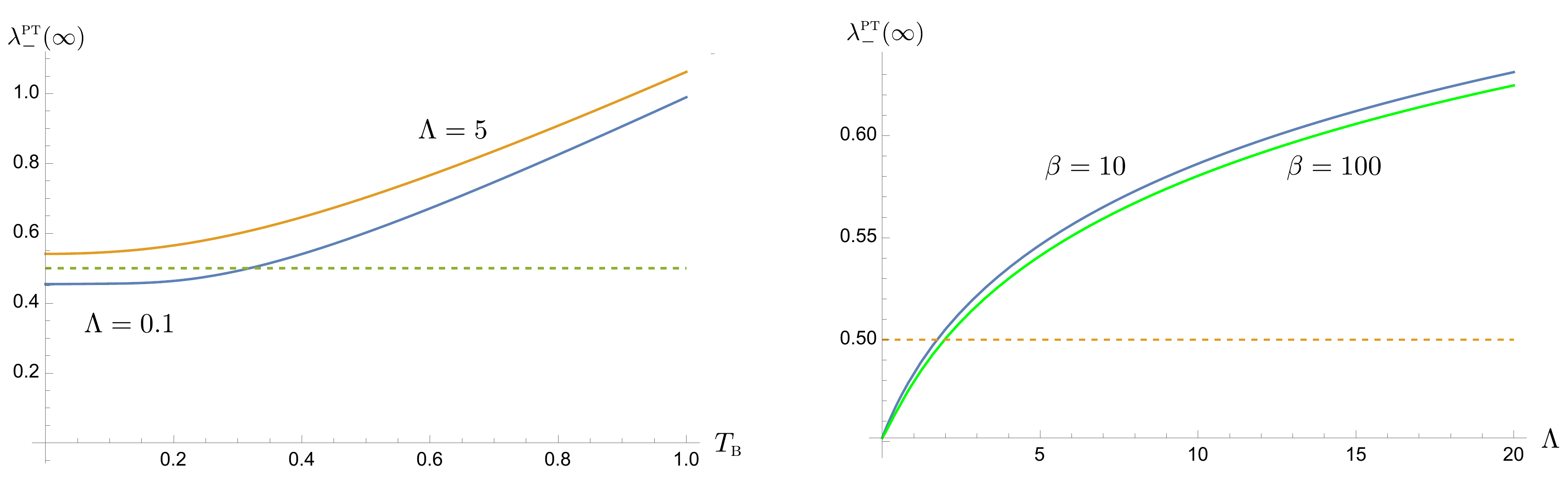}}
    \caption{The late-time behavior of the symplectic eigenvalue $\lambda^{\textsc{pt}}_{-}(\infty)$. The left panel shows its dependence on the bath temperature for two different choices of cutoff scales. The right panel shows the cutoff dependence of $\lambda^{\textsc{pt}}_{-}(\infty)$. Here we choose the parameters, in the unit of $\omega_{\textsc{p}}$, $m=1$, $\gamma=0.5$, and $\sigma=0.2$.}\label{Fi:entLateTime}
\end{figure}

In Fig.~\ref{Fi:entLateTime}, we show the dependence of the symplectic eigenvalue $\lambda^{\textsc{pt}}_{-}$ on the bath temperature and the bath cutoff scale at late times, after the dynamics of the oscillators are fully relaxed. Recall that a large damping constant in the strong oscillator-bath regime is chosen, so the entanglement between the oscillators is typically difficult to maintain at late times. We see a trend, though not very significant, that the longer memory time $\tau_{\textsc{mem}}=\Lambda^{-1}$ or the lower bath temperature $T_{\textsc{b}}=1/\beta$ is prone to keep the oscillators' entanglement at late times. In the left panel, with a longer memory time, the entanglement still exists at higher bath temperatures although this critical temperature does not change much with $\Lambda$ and is still of the order $\mathcal{O}(\omega_{\textsc{p}}^{-1})$. On the other hand, the right panel shows that with a lower bath temperature, the entanglement may still survive for shorter memory times. These examples illustrate the coherent superposition induced by the nonMarkovian memory effect is capable to counter, at least marginally, the debilitating effects accumulated over the whole course of evolution due to the thermal fluctuations.

\section{strongly coupled nonMarkovian dynamics vs weakly coupled Markovian dynamics}\label{S:esbsdfsd}
 In the previous section, we note that the dynamics of a coupled system strongly coupled to  a nonMarkovian bath {seems to behave in some aspects} like that of a system weakly coupled to a Markovian bath, and raised the question whether we can use weakly coupled Markovian linear open systems to approximate strongly coupled nonMarkovian linear open systems? {This issue is of significance because one may argue that even though many open system processes in the real world involve memories and are thus fundamentally nonMarkovian,  we may in practice describe them by using a simpler effectively Markovian formulation. How valid are such prescriptions of convenience? We want to take a closer look at this issue with the help of a concrete example}. 

For the normal modes of the coupled oscillators described by \eqref{E:eotugbsd1}, we suppose there correspond \textit{effective} equations of motion of the normal modes for the weakly coupled Markovian oscillators
\begin{align}
	\ddot{\chi}_{\pm}(t)+2\gamma_{\textsc{eff}}^{(\pm)}\dot{\chi}_{\pm}(t)+\omega_{\pm}^{2}\,\chi_{\pm}(t)+2\gamma_{\textsc{eff}}^{(\pm)}\,\delta(t)\,\chi(0)&=\frac{e}{m}\,\phi_{\pm}(t)\,.\label{E:eotugbsd2}
\end{align}	
For example, when $\omega_{\textsc{p}}=1$, $\gamma=0.5$ and $\Lambda=1$, the effective damping constants of the normal modes are $\gamma^{(+)}_{\textsc{eff}}\simeq0.066\,\omega_{\textsc{p}}$ and $\gamma^{(-)}_{\textsc{eff}}\simeq0.094$. The presence of the delta function $\delta(t)$ in the Markovian case induces an instantaneous kick at the initial time. This gives a $\mathcal{O}(\gamma^{(\pm)}_{\textsc{eff}})$ distortion of $d_1^{(\pm)}(t)$ at time right after the initial time, compare to the case in the absence of the kick. Its effect then diminishes with time. On the other hand, the delta function term does not modify $d_2^{(\pm)}(t)$ because $d_2^{(\pm)}(0)=0$.

\begin{figure}
\centering
    \scalebox{0.37}{\includegraphics{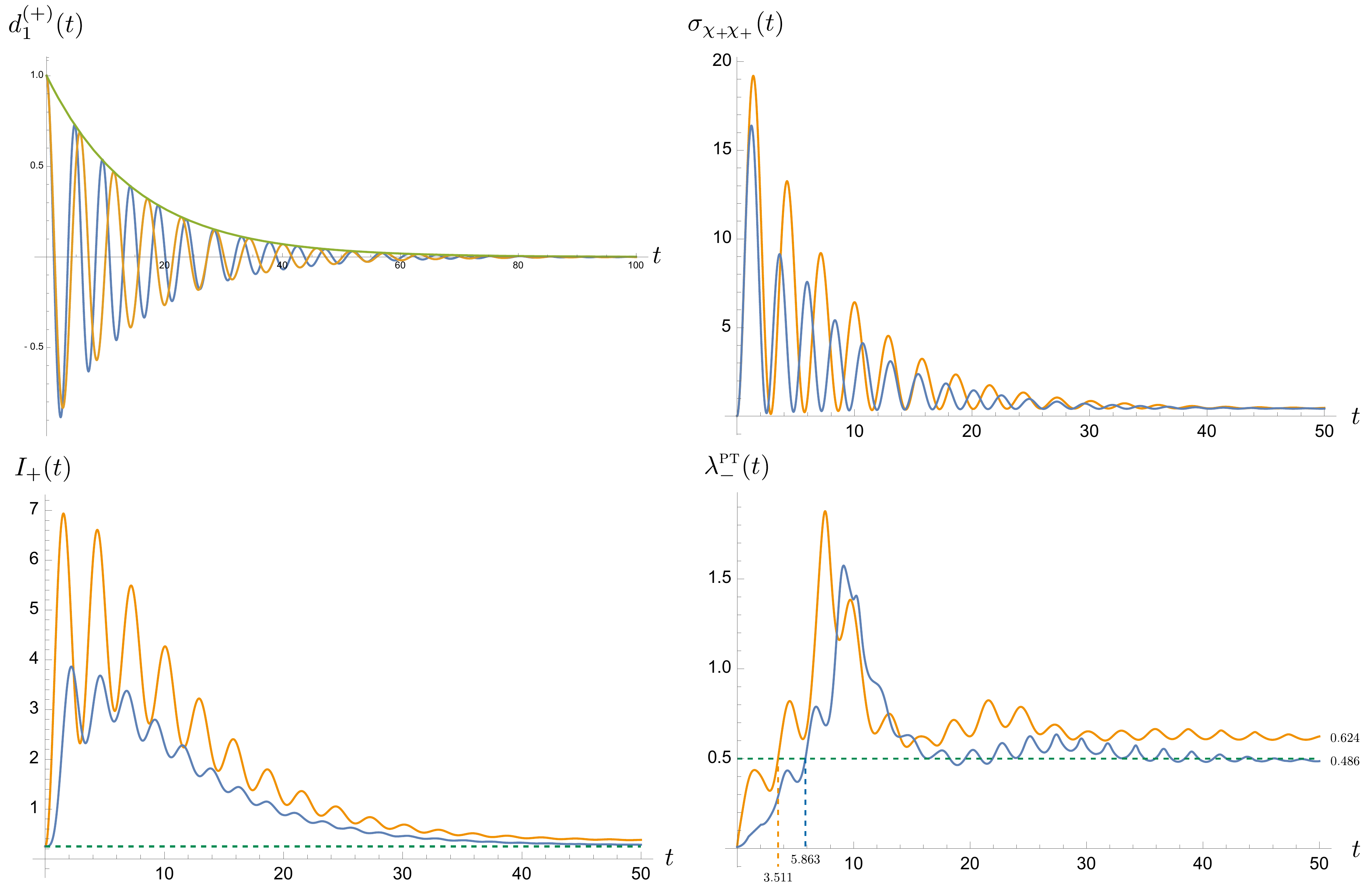}}
    \caption{The time evolution of $d_1^{(+)}(t)$, the covariance matrix element $\sigma_{\chi_+\chi_+}(t)$, the uncertainty relation $I_+(t)$ for the normal mode $+$, and the symplectic eigenvalue $\eta_-^{\textsc{pt}}$. The blue curve corresponding to the nonMarkovian system, \eqref{E:eotugbsd1}, while the orange curve is associated with the effective Markovian system, \eqref{E:eotugbsd2}. For $d_1^{(+)}(t)$, the Green curve describes the envelope of the damped oscillatory evolution. We choose $m=1$, $\omega_{\textsc{p}}=1$, $\gamma=0.5$, $\sigma=0.2$, $\Lambda=1$, and $\beta=10$.}\label{Fi:WMvsSnM}
\end{figure}

Fig.~\ref{Fi:WMvsSnM} shows the time evolution of $d_1^{(+)}(t)$, the covariance matrix element $\sigma_{\chi_+\chi_+}(t)$, the uncertainty relation $I_+(t)$ for the normal mode $+$, and the symplectic eigenvalue $\eta_-^{\textsc{pt}}$. The blue curve corresponds to the strongly coupled nonMarkovian system, \eqref{E:eotugbsd1}, while the orange curve represents the effective, weakly coupled Markovian system, \eqref{E:eotugbsd2}. The fundamental solution in both cases look quite similar. They decay approximately with the same rate, but their phases vary with time. This minor difference starts showing its repercussion effects in the covariance matrix element $\sigma_{\chi_+\chi_+}(t)$, which accounts for the uncertainty of the $\chi_+$ mode, or the degree of coherence. Both curves notably disagree before relaxation. The situation deteriorates even more for $I_+$, the Robertson-Schr\"odinger relation of the mode $+$. Note that both modes are decoupled in the sense of \eqref{E:ktrvsj}. The plot shows that the nonMarkovian description of the same system  seems to have a better control of coherence, compared with the effective Markovian description. Finally when we examine the symplectic eigenvalue $\lambda_-^{\textsc{pt}}$, they give quite a distinct prediction. The original nonMarkovian description shows that the entanglement between the oscillators can be maintained up to $\tau_{\textsc{ent}}\simeq5.863\,\omega_{\textsc{p}}^{-1}$, but the effective Markovian formalism predicts roughly $0.6\,\tau_{\textsc{ent}}$. Moreover, at late times, the original nonMarkovian description shows the presence of residual entanglement in the $\Lambda=1\,\omega_{\textsc{p}}$ and $\beta=10\,\omega_{\textsc{p}}^{-1}$ case, which is in the strong oscillator-bath coupling, low temperature regime. Since the disparity is more than marginal, the Markovian approximation, instead, predicts that the state of the system is separable.

The results in Fig.~\ref{Fi:WMvsSnM}, structured as a hierarchy from the evolutionary phase of the canonical variables, to its dispersion, then to the uncertainty of the constituent of the system, and finally into the integrated correlation among the system, indicate that the approximated Markovian description fails to precisely grasp the phase information embedded in nonMarkovian dynamics of the reduced system of the coupled oscillators. The approximated description tends to give poor predictions of the quantities that involving phases, or coherence. Thus the effective, weak coupling, Markovian description cannot be a sufficiently accurate substitute of the original strong coupling, nonMarkovian, linear open systems, even though the former has extraordinary convenience in computations.

Next we turn to the final issue discussed in this paper: whether the memory effect can be transferred from one party to the other in a bipartite system? 

\section{asymmetric setting}\label{S:egdfd}
Suppose in the system of two coupled oscillators, one of which (oscillator 1) has a finite memory time due to small cutoff scale of its private bath, but the other (oscillator 2) has a negligible memory time due to the large cutoff scale. From the previous discussions, we learn that when they stand alone, oscillator 1 has a much smaller effective damping constant, resulting from the memory effect, in comparison with oscillator 2, so they will relax with different paces. Now when we couple them together, how does the coupled system evolve? Is the evolution dominated by the memoryless system, or by the nonMarkovian system? Or, will the memory effects in oscillator 1 be transferred to oscillator 2, so that it is shared among them, causing the coupled system to evolve in a cooperative way? Further, how does this affect the entanglement dynamics.

Since both private baths have different cutoff scales, the reduced system of coupled oscillators has an asymmetric configuration. We will start from \eqref{E:kgbdfkhdfg}, \eqref{E:bgkdgs}, and \eqref{E:oritjsfg} with 
\begin{equation}
	\tilde{\Gamma}^{(\phi_{i})}(z)=\dfrac{1}{8\pi}\frac{\Lambda_{i}(2\Lambda_{i}+z)}{(\Lambda_{i}+z)^{2}}\,.
\end{equation}
for $i=1$, 2. Since working in the normal modes does not reduce computation hurdles, we will directly compute the covariance matrix elements of the canonical variables of the coupled system. For example, $\sigma_{\chi_1\chi_1}(t)$ will take the form
\begin{align}
    &\sigma_{\chi_1\chi_1}(t)=\bigl[\bm{D}_{1}(t)\bigr]_{11}^{2}\,\langle\chi_{1}^{2}(0)\rangle+2\bigl[\bm{D}_{1}(t)\bigr]_{11}\bigl[\bm{D}_{1}(t)\bigr]_{12}\,\frac{1}{2}\langle\bigl\{\chi_{1}(0),\chi_{2}(0)\bigr\}\rangle+\bigl[\bm{D}_{1}(t)\bigr]_{12}^{2}\,\langle\chi_{2}^{2}(0)\rangle\notag\\
	&\qquad\qquad\qquad+\frac{1}{m^{2}}\,\bigl[\bm{D}_{2}(t)\bigr]_{11}^{2}\,\langle p_{1}^{2}(0)\rangle+\frac{2}{m^{2}}\,\bigl[\bm{D}_{2}(t)\bigr]_{11}\bigl[\bm{D}_{2}(t)\bigr]_{12}\,\frac{1}{2}\langle\bigl\{ p_{1}(0),\,p_{2}(0)\bigr\}\rangle+\frac{1}{m^{2}}\,\bigl[\bm{D}_{2}(t)\bigr]_{12}^{2}\,\langle p_{2}^{2}(0)\rangle\notag\\
	&\qquad\qquad\qquad+\frac{e^{2}}{m^{2}}\int_{0}^{t}\!ds\!\int_{0}^{t}\!ds'\,\Bigl\{\bigl[\bm{D}_{2}(t-s)\bigr]_{11}\bigl[\bm{D}_{2}(t-s')\bigr]_{11}\,G_{H,0}^{(\phi_{1})}(s-s')\Bigr.\notag\\
	&\qquad\qquad\qquad\qquad\qquad\qquad\qquad\qquad\qquad\qquad+\Bigl.\bigl[\bm{D}_{2}(t-s)\bigr]_{12}\bigl[\bm{D}_{2}(t-s')\bigr]_{12}\,G_{H,0}^{(\phi_{2})}(s-s')\Bigr\}\,.
\end{align}
As before, we suppose that the initial state of the coupled oscillator is a two-mode squeezed vacuum state.

\begin{figure}
\centering
    \scalebox{0.35}{\includegraphics{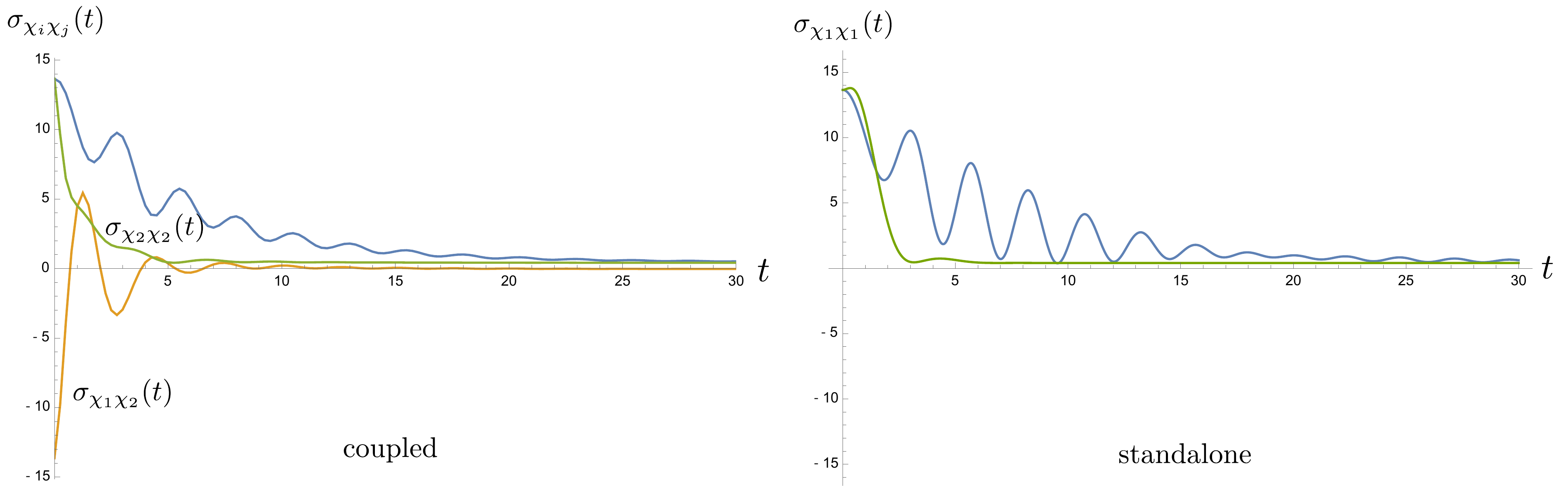}}
    \caption{The time evolution of the covariance matrix element $\sigma_{\chi_i\chi_j}(t)$. The left plot corresponds to the asymmetric setting where the cutoff scales of the private baths are given by $\Lambda_1$ and $\Lambda_2$, respectively. The right plot contains $\sigma_{\chi_1\chi_1}(t)$ for two symmetric settings. The blue curve represents the case both private baths have the same cutoff scale $\Lambda_1$, while the green curve represents the case both private baths have the same cutoff scale $\Lambda_2$. The relevant parameters are $m=1$, $\omega_{\textsc{p}}=1$, $\gamma=0.5$, $\sigma=0.2$, $\beta=10$, $\Lambda_1$=1 and $\Lambda_2=20$, and they are expressed in the unit of $\omega_{\textsc{p}}$.}\label{Fi:asymXX}
\end{figure}

We plot the time evolution of the covariance matrix element $\sigma_{\chi_i\chi_j}(t)$ in Fig.~\ref{Fi:asymXX}. In the right plot, the blue curve corresponds to $\sigma_{\chi_1\chi_1}(t)$ when the private baths of two coupled oscillators have the same cutoff scale $\Lambda_1=1\,\omega_{\textsc{p}}$, while the green curve represents $\sigma_{\chi_1\chi_1}(t)$ when the private baths have the cutoff scale $\Lambda_2=20\,\omega_{\textsc{p}}$. They can be respectively compared to $\sigma_{\chi_1\chi_1}(t)$ and $\sigma_{\chi_2\chi_2}(t)$ in the left plot. For the memoryless case in the right plot (green curve), the covariance matrix element $\sigma_{\chi_1\chi_1}(t)$ decays very fast, and its time evolution is almost fully relaxed when $t\sim4\,\omega_{\textsc{p}}^{-1}$, which is about twice the relaxation time scale $\gamma^{-1}$. On the other hand, in the finite memory case, the blue curve falls off rather slowly. Its oscillatory behavior remains visible when $t=30\,\omega_{\textsc{p}}^{-1}$. The left plot of Fig.~\ref{Fi:asymXX} describes the above asymmetric setting where the oscillator 1's private bath has the cutoff scale $\Lambda_1$, but the oscillator 2's bath has $\Lambda_2$. We immediately see that for such a hybrid system, $\sigma_{\chi_2\chi_2}(t)$ (green curve in the left plot) does not decay as fast as the green curve in the right plot. Nonetheless, $\sigma_{\chi_1\chi_1}(t)$ of oscillator 1 in the hybrid system falls off faster than its counterpart (blue curve) in the right plot. These results imply transfer of the memory effect. Oscillator 2, which is essentially memoryless, benefits from such a transfer because its covariance matrix element shows a prolonged relaxation time (to roughly $t\sim5\,\omega_{\textsc{p}}^{-1}$), but this transfer shortens the memory time of oscillator 1. Thus in the hybrid system, motion of both oscillators in fact takes a cooperative way, and is neither dominated by the memoryless component nor governed by the one with a finite memory. Will their entanglement dynamics shows a similar feature?

\begin{figure}
\centering
    \scalebox{0.4}{\includegraphics{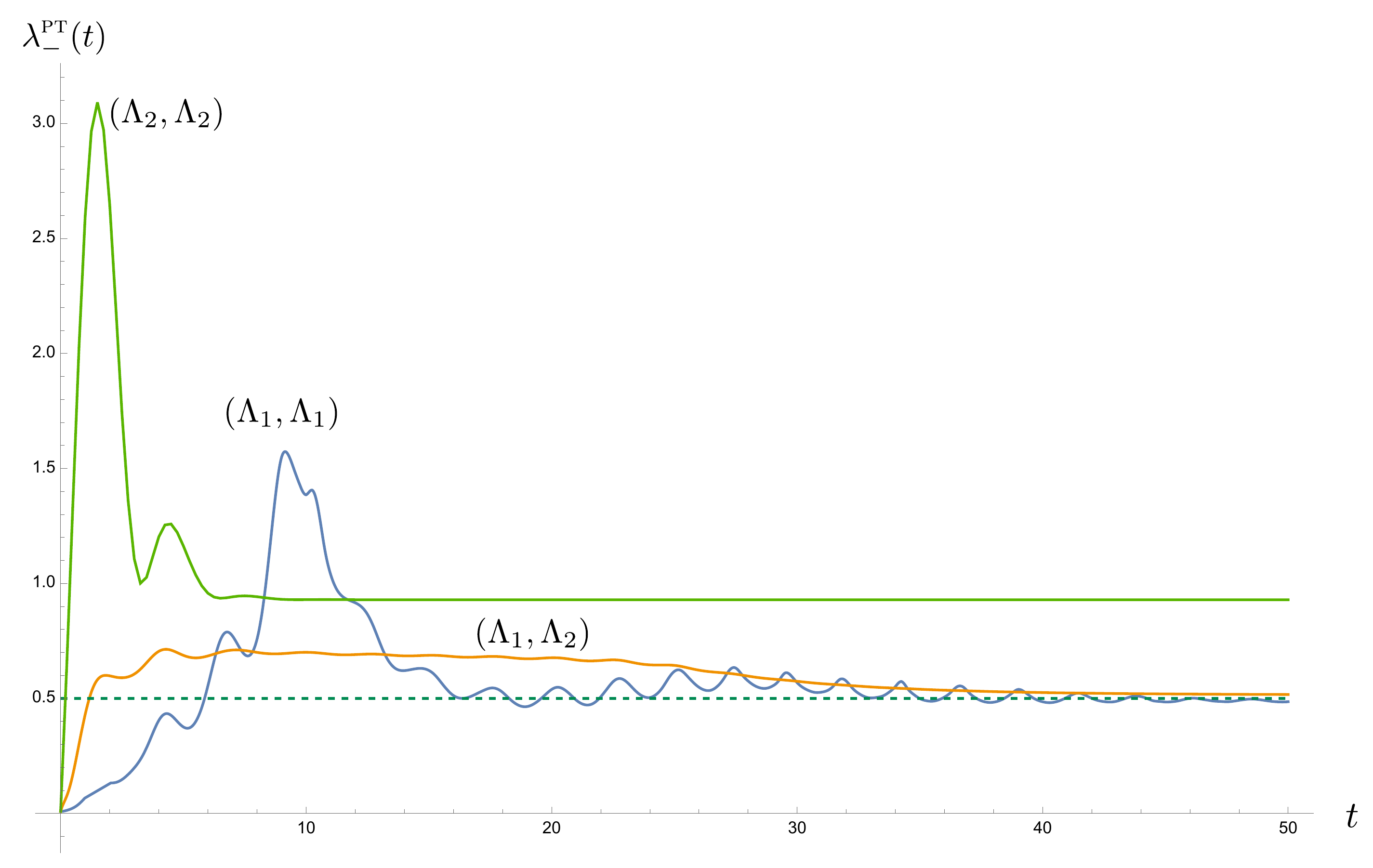}}
    \caption{The time evolution of symplectic eigenvalue $\lambda_-^{\textsc{pt}}(t)$ of the partially transposed covariance matrix $\sigma^{\textsc{pt}}(t)$. The system is entangled when $\lambda_-^{\textsc{pt}}(t)<1/2$. The notation $(\Lambda_i,\Lambda_j)$ denote the cutoff scales of the private baths attached to the coupled oscillators. For example, the asymmetric setting (orange curve) discussed in this section corresponds to the curve with the tag $(\Lambda_1,\Lambda_2)$. Two other symmetric settings $(\Lambda_1,\Lambda_1)$ (blue curve) and $(\Lambda_2,\Lambda_2)$ (green curve) are used to contrast the effect of memory transfer on entanglement dynamics. The relevant parameters are $m=1$, $\omega_{\textsc{p}}=1$, $\gamma=0.5$, $\sigma=0.2$, $\beta=10$, $\Lambda_1$=1 and $\Lambda_2=20$, and they are expressed in the unit of $\omega_{\textsc{p}}$.}\label{Fi:asymSymp}
\end{figure}

Fig.~\ref{Fi:asymSymp} gives the time evolution of the symplectic eigenvalue $\lambda_-^{\textsc{pt}}(t)$ of the partially transposed covariance matrix $\sigma^{\textsc{pt}}(t)$. The asymmetric setting is described by the orange curve, where two private baths have different cutoff scales $\Lambda_1=1\,\omega_{\textsc{p}}$, $\Lambda_2=20\,\omega_{\textsc{p}}$ such that one of the oscillator is Markovian and basically memoryless. The blue curve describes the symmetric setting when both private baths have the same cutoff scale $\Lambda_1$. That is, both oscillators are nonMarkovian and have equal memory time. In contrast to the previous case, we have coupled (almost) Markovian oscillators ($\Lambda_2$) in the green curve. Since the damping constant $\gamma$ is chosen to be 0.5, we find $\lambda_-^{\textsc{pt}}(t)$ associated with the memoryless case settles down to an equilibrium value $0.929$ very quickly, compared to the other two cases. Its value shoots up past $1/2$ when $\tau_{\textsc{ent}}=t\simeq0.197\,\omega_{\textsc{p}}^{-1}$, even though the bath temperature $T_{\textsc{b}}=0.1\,\omega_{\textsc{p}}$, is rather low. In this case, once the state becomes separable, it remains disentangled. For the other extreme when both oscillators have the same, long memory time (blue curve), the initial entanglement is sustained up to $\tau_{\textsc{ent}}\simeq5.863\,\omega_{\textsc{p}}^{-1}$, and at late times the curve falls below the $1/2$ level, whence the entanglement between the oscillators is revived. Finally, in the hybrid case, the duration the system remains entangled happens to fall between two cases of the symmetric setting. We find $\tau_{\textsc{ent}}\simeq1.168\,\omega_{\textsc{p}}^{-1}$, much shorter than the blue curve case but slightly better than the green curve case. At late times, we observe that the curve gradually dips down to 0.571, still above the $1/2$ level, so the system is not entangled at late time. This sloping-down behavior, also seen in the blue curve, may imply the memory effect in one of the private baths is still in effect, because the corresponding bath induces a longer effective relaxation time.

We observe that for the hybrid system during the relaxation regime, the duration the quantum entanglement is sustained over falls much shorter than the case when both oscillators have the same long memory time, and improves slightly compared to the Markovian, memoryless case. A similar phenomenon has been observed in~\cite{hotentanglemt}. There, two coupled oscillators in the {two uncorrelated private bath} setting are attached to their own Markovian private baths, which have different bath temperatures. In that case, we find that if the temperature in one of the private baths is raised beyond the critical temperature of the order $\omega_{\textsc{p}}^{-1}$, then the entanglement between the oscillator is lost even though the other bath is kept at temperature much lower than the critical value. Similarly, here since one of the oscillator is memoryless, the coupled system formed purely by this oscillator is supposed to have a short entanglement time $\tau_{\textsc{ent}}$ (the green curve). Thus even it is coupled to a nonMarkovian oscillator, as in the hybrid case, the entanglement duration still fails to gain any improvement. That is, the entanglement duration of a bipartite system is predominantly controlled by the component that least favors the sustainability of system entanglement.

\section{Discussion}\label{S:gdiser}

In this paper we present a detailed analysis of the effects of nonMarkovian dynamics on the entanglement dynamics between two coupled oscillators, each of which has its own private bath. In particular, we are interested in how the memory effects in the baths affect  the full course of entanglement evolution -- from the initial time to the moment the reduced system relaxes to the equilibrium state. We break down the analysis in parts to examine the imprints the nonMarkovian baths leave  on the solutions of Langevin equation, time evolution of the covariance matrix, and  on the negativity, serving as our entanglement measure.   Based on the results we obtained we now can answer the questions posed in the Introduction in the following: 
\begin{enumerate}
    \item How does the nonMarkovianity in the bath affect the system's entanglement ?
    \item Which time regime does the bath's nonMarkovianity benefit the system's entanglement most?
    \item Can the memory of one nonMarkovian bath be passed on to another Markovian bath?
    \item Does this memory transfer help to  sustain  the system's entanglement dynamics?
\end{enumerate}

{
\paragraph{\textit{How does the nonMarkovianity in the bath affect the system's entanglement? --}} This is most easily seen when both oscillator-private bath pairs have the same setting. For the homogeneous solution to the Langevin equations, Fig.~\ref{Fi:effDamp} shows that the solution damps at a slower rate, determined by the cutoff scale $\Lambda$ in the bath's spectral density than what would be given by the damping constant $\gamma$. In this paper it is chosen in the strong coupling regime.  For the range of chosen parameters, the actual damping rate can be well  approximated  by the effective damping constant $\gamma_{\textsc{eff}}$ in \eqref{E:fgskd}. Roughly speaking, a smaller cutoff scale leads to a lower value of the effective damping constant. Thus the original system that is strongly coupled to the nonMarkovian thermal bath seems to behave like a system weakly coupled to a Markovian bath. The phenomenon of the weakened damping may be understood by comparing the memory time and the dynamical time scale of motion. When the memory time is much longer than the period of the oscillator, the contributions from many previous cycles  coherently add  up to the current state of motion, and thus compensate for the attenuation caused by damping.  Alternatively,  a cutoff scale, especially a scale  smaller than the resonance frequency of the oscillator driven by the quantum fluctuations of the bath, implies that the oscillator does not lose its energy back to the bath as effectively. This characteristic of a weakened damping is likewise passed on to the time evolution of the covariance matrix elements and the entanglement measure, as shown in Figs.~\ref{Fi:coVmemoryTemp} and Fig.~\ref{Fi:enTmemoryTemp}. These two plots, as well as, Figs.~\ref{Fi:entLTvsLamb} and \ref{Fi:entLateTime} show that the nonMarkovian effect becomes inconsequential at higher bath temperature.}

\paragraph{\textit{Which time regime does the nonMarkovianity in the bath benefit the system's entanglement most? --}} Coherent superposition of the history of the system's dynamics effectively reduces the damping, and prolongs the relaxation time scale. However, as the system approaches relaxation, its dynamics is gradually taken over by its private thermal bath. The role of the initial conditions is exponentially suppressed in this case. The prolonged relaxation thus lead to the extended intervention of the thermal fluctuations of the bath. This accumulation seems to almost cancel out the benefit from the memory effect. This can be seen from Figs.~\ref{Fi:coVmemoryTemp}. These observations are reflected in the time evolution of entanglement in Fig.~\ref{Fi:enTmemoryTemp}. Suppose that the oscillators are initially entangled, we see that in the intermediate time range, the duration of entanglement is proportional to the memory time, and it lasts about $1/3$ of the relaxation time, but at late times when the dynamics reaches the steady state, the value of the symplectic eigenvalue $\lambda_-^{\textsc{pt}}$ of the partially transposed covariance matrix barely benefit from the bath nonMarkovianity, shown in Fig.~\ref{Fi:entLateTime}. Thus {\it bath nonMarkovianity has a limited capacity to improve the late-time entanglement.}

{
\paragraph{\textit{Can the memory of one nonMarkovian bath be passed on to another Markovian bath? --}} Note this question is not about whether the memory can be faithfully mapped from one party to the other. Rather, we ask whether a system with a shorter memory time can be inducted to a longer memory time when it is coupled to another system that has a long memory time. As far as the dynamics is concerned, the answer seems to be yes. In Fig.~\ref{Fi:asymXX}, where oscillator 2, which is (almost) Markovian, is coupled to oscillator 1, which possess a long memory time. The results in the right panel of Fig.~\ref{Fi:asymXX} serves as a contrast, showing the time evolution of the position uncertainty when both oscillators are oscillator 1 (blue curve) or both oscillators are oscillator 2 (green curve). We see that the green curve falls to a constant very fast within the time scale $\gamma^{-1}$. The green curve on the left panel of Fig.~\ref{Fi:asymXX} on the other hand shows the counterpart when oscillator 2 is coupled to oscillator 1. It does not decay as fast as the former, indicating that in this case the effective damping is weaker, or in other words, the ``effective'' memory time becomes longer. In this sense, the memory of oscillator 2 indeed improves. But, there is a catch. The effective memory time of oscillator 1 deteriorates when it is coupled to oscillator 2, compared to the case when it is coupled to an identical copy (comparing the blue curves in both panels of Fig.~\ref{Fi:asymXX}). Therefore, {\it a system with a short memory time can acquire improvement when it is coupled to another system with a long memory time, but, at the cost of the latter. } 
}

{
\paragraph{\textit{Does this memory transfer help to  sustain the system's entanglement dynamics? --}} when systems of different memory times are coupled together, from the conclusion brought forth above, we note that the two oscillators still have quite distinct time scales in their dynamics, even though one has improved a bit and the other degraded a bit. The consequence, as shown in Fig.~\ref{Fi:asymSymp}, is that the entanglement time is improved, compared to the case when both oscillators are oscillator 2, but is shortened, compared to the other extreme. Overall, our results show {\it the sustainability of the bipartite entanglement is determined by the party which breaks off entanglement most easily}.
}

{
One more question arises when we delineate the role of bath nonMarkovianity. Earlier,  we note that the solution to the equation of motion for a system strongly coupled to a nonMarkovian system behaves similar to that for a system weakly coupled to a Markovian bath in the sense that they can have the same decaying rate with minor difference in the phases (shown in top left plot in Fig.~\ref{Fi:WMvsSnM}). We may ask whether we can effectively use the latter to approximate the former because the numerical computation of a Markovian system is much less demanding, and the theoretical analysis is simpler? Or pushing it to the conceptual extreme, is it possible that the system we thought to be Markovian is actually a mere mirage of a nonMarkovian system with stronger coupling? In either case, it is then important to examine the similarity and disparity of both descriptions. The additional plots in Fig.~\ref{Fi:WMvsSnM} show that the minor phase difference in the solution $d_1^{(+)}(t)$ results in broader dissimilarity in the dispersion, uncertainty relation and then bipartite entanglement. This reflects subtleties of the phase carried in a nonMarkovian system. We assert that {\it the effective Markovian description is not sufficiently accurate to capture the needed phase information, so it cannot faithfully mimic a nonMarkovian system}.
} \\

\noindent{
\bf Acknowledgment} J.-T. Hsiang is supported by the Ministry of Science and Technology of Taiwan, R.O.C. under Grant No.~MOST 110-2811-M-008-522. O. Ar{\i}soy is supported in part by the Maryland Center for Fundamental Physics.

\newpage
\appendix
\section{positive partial transpose criterion and negativity}\label{S:jgvsf}
Here is the concise summary of negativity for the two-mode Gaussian states. Let $\bm{R}=(\chi_{1}, p_{1}, \chi_{2}, p_{2})^{T}$ accommodates the canonical variables of two Gaussian modes, and we introduce the covariance matrix $\bm{\sigma}$ by
\begin{equation}
	\bm{\sigma}=\frac{1}{2}\langle\bigl\{\bm{R},\bm{R}^{T}\bigr\}\rangle-\langle\bm{R}\rangle\cdot\langle\bm{R}^{T}\rangle=\begin{pmatrix}\bm{A}&\bm{C}\\\bm{C}^{T}&\bm{B}\end{pmatrix}\,,
\end{equation}
with $\bm{\Omega}^{-1}=\bm{\Omega}^{T}=-\bm{\Omega}$, and
\begin{align}
	\bm{\Omega}&=\begin{pmatrix}\bm{J} &\bm{0}\\\bm{0}&\bm{J}\end{pmatrix}\,,&\bm{J}&=\begin{pmatrix}0 &+1\\-1&0\end{pmatrix}\,,
\end{align}
such that the canonical commutation relation takes the form
\begin{equation}
	\bigl[\bm{R},\bm{R}^{T}\bigr]=i\,\bm{\Omega}\,,
\end{equation}
and the uncertainty principle can be re-formulated as~\cite{simon}
\begin{equation}
	\bm{\sigma}+\frac{i}{2}\,\bm{\Omega}\geq0\,.\label{E:hrvgdjf}
\end{equation}
The $\mathrm{Sp}(2,\mathbb{R})\otimes\mathrm{Sp}(2,\mathbb{R})$ invariants associated with $\bm{\sigma}$ are
\begin{align}
	I_{1}&=\det\bm{A}\,,&I_{2}&=\det\bm{B}\,,&I_{3}&=\det\bm{C}\,,&I_{4}&=\operatorname{Tr}\bigl\{\bm{A}\cdot\bm{J}\cdot\bm{C}\cdot\bm{J}\cdot\bm{B}\cdot\bm{J}\cdot\bm{C}^{T}\cdot\bm{J}\bigr\}\,,
\end{align}
with $\det\bm{\sigma}=I_{1}I_{2}+I_{3}^{2}-I_{4}$. The uncertainty principle \eqref{E:hrvgdjf} can be cast into an $\mathrm{Sp}(2,\mathbb{R})\otimes\mathrm{Sp}(2,\mathbb{R})$-invariant statement
\begin{equation}
	I_{1}+I_{2}+2I_{3}\leq\frac{1}{4}+4\det\bm{\sigma}\,.
\end{equation}
It is then convenient to introduce another invariant $\Delta(\bm{\sigma})=I_{1}+I_{2}+2I_{3}$.

According to Williamson's theorem, we can find a suitable symplectic transformation $\bm{S}$ to diagonalize the covariance matrix $\bm{\sigma}$	
\begin{align}\label{E:ktvkgsf}
	\bm{\sigma}&=\bm{S}^{T}\cdot\bm{K}\cdot\bm{S}\,,&\bm{K}&=\begin{pmatrix}\lambda_{-}\bm{I}_{2} &\bm{0}\\\bm{0} &\lambda_{+}\bm{I}_{2}\end{pmatrix}\,,&\bm{I}_{2}&=\begin{pmatrix}1 &0\\0 &1\end{pmatrix}\,,
\end{align}
where $\lambda_{\pm}$ are the symplectic eigenvalues of $\bm{\sigma}$ with $\lambda_{-}\leq\lambda_{+}$. Alternatively, we can find a suitable unitary transformation $\bm{V}$ to diagonalized $i\,\bm{\Omega}\cdot\bm{\sigma}$
\begin{equation}
	i\,\bm{\Omega}\cdot\bm{\sigma}=\bm{V}^{-1}\cdot\operatorname{diag}\bigl(+\lambda_{-},-\lambda_{-},+\lambda_{+},-\lambda_{+}\bigr)\cdot\bm{V}\,,
\end{equation}
such that $\pm\lambda_{\pm}$ are the ordinary eigenvalues of $i\,\bm{\Omega}\cdot\bm{\sigma}$.

By definition of the symplectic transformation $\bm{\Omega}=\bm{S}^{T}\cdot\bm{\Omega}\cdot\bm{S}$, we have $\det\bm{S}=1$, so that
\begin{equation}
	\det\bm{\sigma}=\det\bm{K}=\lambda_{-}^{2}\lambda_{+}^{2}\,.
\end{equation}
On the other hand, since $\Delta(\bm{\sigma})$ is an $\mathrm{Sp}(2,\mathbb{R})\otimes\mathrm{Sp}(2,\mathbb{R})$ invariant, we have 
\begin{equation}
	\Delta(\bm{\sigma})=\lambda_{-}^{2}+\lambda_{+}^{2}\,.
\end{equation}
Together we arrive at an expression of the symplectic eigenvalues of $\bm{\sigma}$ in terms of two invariants $\Delta(\bm{\sigma})$ and $\det\bm{\sigma}$
\begin{equation}
	\lambda^{2}_{\pm}=\frac{1}{2}\Bigl[\Delta(\bm{\sigma})\pm\sqrt{\Delta^{2}(\bm{\sigma})-4\det\bm{\sigma}}\Bigr]\,,
\end{equation}
for the covariance matrix $\bm{\sigma}$. Eq.~\eqref{E:ktvkgsf} implies that the uncertainty principle becomes
\begin{align}
	\bm{K}+\frac{i}{2}\,\bm{\Omega}&\geq0\,,&&\Rightarrow&\bigl(\lambda_{+}^{2}-\frac{1}{4}\bigr)\bigl(\lambda_{-}^{2}-\frac{1}{4}\bigr)&\geq0\,,&&\Rightarrow&\lambda_{\pm}&\geq\frac{1}{2}\,.
\end{align}
The symplectic eigenvalues $\lambda_{\pm}$ are positive because $\bm{\sigma}$ is a positive-definite, symmetric real matrix.

For the two-mode Gaussian system, we may write the covariance matrix into a canonical form~\cite{duan}
\begin{equation}\label{E:ajgjhsdfw}
	\bm{\sigma}=\begin{pmatrix} a &0 &c &0\\0 &a &0 &d\\c &0 &b &0\\0 &d &0 &b\end{pmatrix}\,,
\end{equation}
via suitable local symplectic transformation $\operatorname{Sp}(2,\mathbb{R})\otimes\operatorname{Sp}(2,\mathbb{R})$, that is, local changes of the canonical variables. This allows us to readily find the $\operatorname{Sp}(2,\mathbb{R})\otimes\operatorname{Sp}(2,\mathbb{R})$ invariants,
Then we have
\begin{align}
	I_{1}&=a^{2}\,,&I_{2}&=b^{2}\,,&I_{3}&=cd\,,&I_{4}&=ab\bigl(c^{2}+d^{2}\bigr)\,,\\
 \det\bm{\sigma}&=\bigl(ab-c^{2}\bigr)\bigl(ab-d^{2}\bigr)\,,&&&\Delta(\bm{\sigma})&=a^{2}+b^{2}+2cd\,.
\end{align}

For a bipartite system $AB$, if the state $\rho$ is separable, then the partial transpose of $\rho$ is still non-negative, and vice versa, a bona fide density matrix~\cite{ppt}. We have a corresponding uncertainty relation,
\begin{equation}
	\bm{\sigma}^{\textsc{pt}}+\frac{i}{2}\,\bm{\Omega}\geq0\,.\label{E:chvbxjhdfs}
\end{equation}
The covariance matrix $\bm{\sigma}^{\textsc{pt}}$ is the partial transpose of the original covariance matrix via $\bm{\sigma}^{\textsc{pt}}=\bm{\Lambda}\cdot\bm{\sigma}\cdot\bm{\Lambda}$. When the transposition is carried out with respect to $B$, the matrix $\bm{\Lambda}$ has the form $\bm{\Lambda}=\operatorname{diag}(+1,+1,+1,-1)$, and then $\bm{\sigma}^{\textsc{pt}}$ has the form
\begin{equation}
	\bm{\sigma}^{\textsc{pt}}=\begin{pmatrix}\bm{A}&\bm{C}'\\\bm{C}'^{T}&\bm{B}\end{pmatrix}\,,
\end{equation}
with $\det\bm{C}'=-\det\bm{C}$. Note that $\bm{\Lambda}$ is not a symplectic matrix, that is, $\bm{\Lambda}\cdot\bm{\Omega}\cdot\bm{\Lambda}\neq\bm{\Omega}$.

We can likewise write down the corresponding $\mathrm{Sp}(2,\mathbb{R})\otimes\mathrm{Sp}(2,\mathbb{R})$ invariants, denoted by an extra prime, for the partially transposed covariance matrix. Since $\bm{\Lambda}^{2}=\bm{I}$, we find that they are related to the invariants associated with $\bm{\sigma}$ by
\begin{align}
	I'_{3}&=-I_{3}\,,&I'_{4}&=I_{4}\,,&\det\bm{\sigma}^{\textsc{pt}}&=\det\bm{\sigma}\,, &\Delta(\bm{\sigma}^{\textsc{pt}})&=I_{1}+I_{2}-2I_{3}\,, 
\end{align}
and the corresponding uncertainty relation is given by
\begin{equation}
	I_{1}I_{2}+\bigl(\frac{1}{4}+I_{3}\bigr)^{2}-I_{4}\geq\frac{1}{4}\bigl(I_{1}+I_{2}\bigr)\,.\label{E:kgbdf}
\end{equation}
Since we arrive at this expression by assuming the state $\rho$ is separable, Eq.~\eqref{E:kgbdf} serves as the necessary criterion of separability. Following the earlier arguments, we find it implies
\begin{equation}
	\lambda^{\textsc{pt}}_{\pm}\geq\frac{1}{2}\,,
\end{equation}
where
\begin{equation}\label{E:rgtetsd}
	(\lambda^{\textsc{pt}}_{\pm})^{2}=\frac{1}{2}\Bigl[\Delta(\bm{\sigma}^{\textsc{pt}})\pm\sqrt{\Delta^{2}(\bm{\sigma}^{\textsc{pt}})-4\det\bm{\sigma}^{\textsc{pt}}}\Bigr]\,.
\end{equation}
The criterion \eqref{E:kgbdf} happens to be sufficient for the two-mode Gaussian state. Thus when $\lambda_-^{\textsc{pt}}<1/2$, the state is not separable, that is, entangled.

This allows us to introduce a quantifiable entanglement measure~\cite{virmani,adesso05}, logarithmic negativity, $\mathcal{E}_N$ by~\cite{negativity}
\begin{equation}
    \mathcal{E}_N=\max\{0,-\ln(2\lambda_{-}^{\textsc{pt}})\}\,.
\end{equation}            
The Gaussian state is entangled when $\mathcal{E}_N>0$.

\section{two-mode squeezed state}\label{S:kgbdf}
The two-mode squeezed thermal state $\rho_{\textsc{tmsq}}^{(\beta)}$ is defined by
\begin{equation}
	\rho_{\textsc{tmsq}}^{(\beta)}=\mathcal{S}_{2}^{\vphantom{\dagger}}\,\rho_{\beta}\,\mathcal{S}_{2}^{\dagger}\,.
\end{equation}
The operator $\mathcal{S}_{2}$ is the two-mode squeeze operator $\mathcal{S}_{2}=\exp\Bigl[\zeta^{*}a_{1}^{\vphantom{\dagger}}a_{2}^{\vphantom{\dagger}}-\zeta\,a_{1}^{\dagger}a_{2}^{\dagger}\Bigr]$ with $\zeta=\eta\,e^{i\theta}$, $\eta\geq0$ and $0\leq\theta<2\pi$. Its action on the annihilation operator, say $a_1$ of mode 1, gives
\begin{equation}
	\mathcal{S}_{2}^{\dagger}\,a_{1}^{\vphantom{\dagger}}\mathcal{S}_{2}^{\vphantom{\dagger}}=\cosh\eta\,a_{1}-e^{i\theta}\sinh\eta\,a_{2}^{\dagger}\,.
\end{equation}
Note that the two-mode squeeze operator is symmetric in $a_{1}$ and $a_{2}$, so a similar result for $a_{2}$ can be obtained with $1\leftrightarrow2$. The annihilation operators, $a_1$ and $a_2$, of mode 1 and mode 2 satisfy the canonical commutation relation $[a_j^{\vphantom{\dagger}},a_k^{\dagger}]=\delta_{jk}$ with $i$, $j=1$, 2. The nontrivial moments of $a_i$ in the two-mode squeezed thermal state are 
\begin{align}
	\langle a_{i}^{2}\rangle_{\textsc{tmsq}}^{(\beta)}&=\operatorname{Tr}\Bigl\{\rho_{\beta}\,\mathcal{S}_{2}^{\dagger}\,a_{i}^{2}\,\mathcal{S}_{2}^{\vphantom{\dagger}}\Bigr\}=0\,,&\langle a_{i}^{\dagger2}\rangle_{\textsc{tmsq}}^{(\beta)}&=0\,,\label{E:ghvsjg1}\\
	\langle a_{1}^{\vphantom{\dagger}}a_{1}^{\dagger}\rangle_{\textsc{tmsq}}^{(\beta)}&=\bigl(\bar{n}_{1}+1\bigr)\,\cosh^{2}\eta+\bar{n}_{2}\,\sinh^{2}\eta\,,&\langle a_{1}^{\dagger}a_{1}^{\vphantom{\dagger}}\rangle_{\textsc{tmsq}}^{(\beta)}&=\bar{n}_{1}\,\cosh^{2}\eta+\bigl(\bar{n}_{2}+1\bigr)\,\sinh^{2}\eta\,,\\
	\langle a_{2}^{\vphantom{\dagger}}a_{2}^{\dagger}\rangle_{\textsc{tmsq}}^{(\beta)}&=\bigl(\bar{n}_{2}+1\bigr)\,\cosh^{2}\eta+\bar{n}_{1}\,\sinh^{2}\eta\,,&\langle a_{2}^{\dagger}a_{2}^{\vphantom{\dagger}}\rangle_{\textsc{tmsq}}^{(\beta)}&=\bar{n}_{2}\,\cosh^{2}\eta+\bigl(\bar{n}_{1}+1\bigr)\,\sinh^{2}\eta\,,\\
	\langle a_{1}a_{2}\rangle_{\textsc{tmsq}}^{(\beta)}&=-\frac{e^{+i\theta}}{2}\bigl(\bar{n}_{1}+\bar{n}_{2}+1\bigr)\,\sinh2\eta\,,&\langle a_{1}^{\dagger}a_{2}^{\dagger}\rangle_{\textsc{tmsq}}^{(\beta)}&=-\frac{e^{-i\theta}}{2}\bigl(\bar{n}_{1}+\bar{n}_{2}+1\bigr)\,\sinh2\eta\,,\label{E:ghvsjg4}
\end{align}
where $\bar{n}_i$ is the average particle number density in the thermal state $\rho_{\beta}$.

Now given the canonical pair $(\chi_j,p_k)$ with $[\chi_j,p_k]=i\,\delta_{jk}$, which are expanded by $a_i^{\vphantom{\dagger}}$, $a_i^{\dagger}$ via
\begin{align}
	\chi_{i}&=\frac{1}{\sqrt{2m\omega}}\bigl(a_{i}^{\dagger}+a_{i}^{\vphantom{\dagger}}\bigr)\,, &p_{i}&=i\sqrt{\frac{m\omega}{2}}\bigl(a_{i}^{\dagger}-a_{i}^{\vphantom{\dagger}}\bigr)\,,
\end{align}
we find the corresponding covariance matrix elements given by
\begin{align}
	\langle\chi_{1}^{2}\rangle&=\frac{1}{m\omega}\,\Bigl[\bigl(\bar{n}_{1}+\frac{1}{2}\bigr)\,\cosh^{2}\eta+\bigl(\bar{n}_{2}+\frac{1}{2}\bigr)\,\sinh^{2}\eta\Bigr]\,,\label{E:bsgvsj1}\\
	\langle\chi_{2}^{2}\rangle&=\frac{1}{m\omega}\,\Bigl[\bigl(\bar{n}_{2}+\frac{1}{2}\bigr)\,\cosh^{2}\eta+\bigl(\bar{n}_{1}+\frac{1}{2}\bigr)\,\sinh^{2}\eta\Bigr]\,,\\
	\langle p_{1}^{2}\rangle&=m\omega\,\Bigl[\bigl(\bar{n}_{1}+\frac{1}{2}\bigr)\,\cosh^{2}\eta+\bigl(\bar{n}_{2}+\frac{1}{2}\bigr)\,\sinh^{2}\eta\Bigr]\,,\\
	\langle p_{2}^{2}\rangle&=m\omega\,\Bigl[\bigl(\bar{n}_{2}+\frac{1}{2}\bigr)\,\cosh^{2}\eta+\bigl(\bar{n}_{1}+\frac{1}{2}\bigr)\,\sinh^{2}\eta\Bigr]\,,\\
	\frac{1}{2}\,\langle\bigl\{\chi_{1},\,\chi_{2}\bigr\}\rangle&=-\frac{1}{2m\omega}\,\bigl(\bar{n}_{1}+\bar{n}_{2}+1\bigr)\,\sinh2\eta\,\cos\theta\,,\\
	\frac{1}{2}\,\langle\bigl\{p_{1},\,p_{2}\bigr\}\rangle&=+\frac{m\omega}{2}\,\bigl(\bar{n}_{1}+\bar{n}_{2}+1\bigr)\,\sinh2\eta\,\cos\theta\,,\\
	\frac{1}{2}\,\langle\bigl\{\chi_{1},\,p_{2}\bigr\}\rangle&=-\frac{1}{2}\,\bigl(\bar{n}_{1}+\bar{n}_{2}+1\bigr)\,\sinh2\eta\,\sin\theta\,,\\
	\frac{1}{2}\,\langle\bigl\{\chi_{2},\,p_{1}\bigr\}\rangle&=-\frac{1}{2}\,\bigl(\bar{n}_{1}+\bar{n}_{2}+1\bigr)\,\sinh2\eta\,\sin\theta\,,\\
	\frac{1}{2}\,\langle\bigl\{\chi_{1},\,p_{1}\bigr\}\rangle&=0\,,\\
	\frac{1}{2}\,\langle\bigl\{\chi_{2},\,p_{2}\bigr\}\rangle&=0\,.\label{E:bsgvsj10}
\end{align}
Thus the symplectic eigenvalues $\lambda_{\pm}$ are given by
\begin{equation}\label{E:kgbdvhse}
	\lambda_{\pm}=\bar{n}_{i}+\frac{1}{2}\,,
\end{equation}
as expected. We also note that if we choose $\theta=0$ and $m\omega=1$, then the covariance matrix of the two-mode squeezed thermal state will take the canonical form \eqref{E:ajgjhsdfw}.

Then from Appendix~\ref{S:jgvsf}, the $\mathrm{Sp}(2,\mathbb{R})\otimes\mathrm{Sp}(2,\mathbb{R})$ invariants associated with covariance matrix $\bm{\sigma}$,~\eqref{E:bsgvsj1}--\eqref{E:bsgvsj10}, are
\begin{align}
	I_{1}&=\Bigl[\bigl(\bar{n}_{1}+\frac{1}{2}\bigr)\,\cosh^{2}\eta+\bigl(\bar{n}_{2}+\frac{1}{2}\bigr)\,\sinh^{2}\eta\Bigr]^{2}\,,\\
	I_{2}&=\Bigl[\bigl(\bar{n}_{2}+\frac{1}{2}\bigr)\,\cosh^{2}\eta+\bigl(\bar{n}_{1}+\frac{1}{2}\bigr)\,\sinh^{2}\eta\Bigr]^{2}\,,\\
	I_{3}&=-\frac{1}{4}\,\bigl(\bar{n}_{1}+\bar{n}_{2}+1\bigr)^{2}\sinh^{2}2\eta\,,\\
	I_{4}&=\frac{1}{16}\,\bigl(\bar{n}_{1}+\bar{n}_{2}+1\bigr)^{2}\Bigl[\bigl(1+2\bar{n}_{1}\bigr)\bigl(1+2\bar{n}_{2}\bigr)-\bigl(\bar{n}_{1}-\bar{n}_{2}\bigr)^{2}+\cosh4\eta\,\bigl(\bar{n}_{1}+\bar{n}_{2}+1\bigr)^{2}\Bigr]\sinh^{2}2\eta\,,\\
	\det\bm{\sigma}&=\bigl(\bar{n}_{1}+\frac{1}{2}\bigr)^{2}\bigl(\bar{n}_{2}+\frac{1}{2}\bigr)^{2}\,,\\
	\Delta(\bm{\sigma})&=\frac{1}{2}+\bar{n}_{1}+\bar{n}_{2}+\bar{n}_{1}^{2}+\bar{n}_{2}^{2}\,,\\
	\Delta(\bm{\sigma}^{\textsc{pt}})&=\frac{1}{2}\,\bigl(\bar{n}_{1}-\bar{n}_{2}\bigr)^{2}+\frac{1}{2}\,\bigl(\bar{n}_{1}+\bar{n}_{2}+1\bigr)^{2}\cosh4\eta\,.
\end{align}
By \eqref{E:rgtetsd}, these invariant allow us to obtain the symplectic eigenvalues
\begin{align}
	\bigl(\lambda^{\textsc{pt}}_{\pm}\bigr)^{2}&=\frac{1}{4}\Bigl[\bigl(\bar{n}_{1}-\bar{n}_{2}\bigr)^{2}+\bigl(\bar{n}_{1}+\bar{n}_{2}+1\bigr)^{2}\cosh4\eta\Bigr]\notag\\
	&\qquad\qquad\qquad\pm\frac{1}{4}\sqrt{\bigl[\bigl(\bar{n}_{1}-\bar{n}_{2}\bigr)^{2}+\bigl(\bar{n}_{1}+\bar{n}_{2}+1\bigr)^{2}\cosh4\eta\bigr]^{2}-\bigl(1+2\bar{n}_{1}\bigr)^{2}\bigl(1+2\bar{n}_{2}\bigr)^{2}}\,.
\end{align}
In the special case $\bar{n}_{1}=\bar{n}_{2}=\bar{n}$, we have
\begin{equation}\label{E:kgfbdf}
	\lambda^{\textsc{pt}}_{\pm}=e^{\pm2\eta}\,\bigl(\bar{n}+\frac{1}{2}\bigr)\,.
\end{equation}
This expression is of particular significance. It tells that in the two mode squeezed thermal state, squeezing and thermal fluctuations play competing roles in sustaining quantum entanglement. The effect of thermal fluctuations are reflected in the mean occupation number of the mode, which will grows proportional to the bath temperature in the high temperature limit because stronger thermal fluctuations tend to excite the mode to higher levels. Furthermore, Eq.~\eqref{E:kgfbdf} indicates that even at high temperature regime, the two modes are still able to maintain entanglement when the squeezing is sufficiently large, compared with the contribution from the thermal fluctuations. That is, since the entanglement occurs when $\lambda^{\textsc{pt}}_{-}<1/2$, we find that it is possible only when the squeezing parameter is at least as large as
\begin{align}
	e^{-2\eta}\,\bigl(\bar{n}+\frac{1}{2}\bigr)&<\frac{1}{2}\,,&&\Rightarrow&\eta&>\frac{1}{2}\,\ln\bigl(2\bar{n}+1\bigr)=\frac{1}{2}\,\ln\coth\frac{\beta\omega}{2}\,.
\end{align}
In the high temperature limit, the lower bound of the squeeze parameter $\eta$ to generate entanglement is
\begin{equation}
    \eta>\frac{T_{\textsc{b}}}{\omega}\,,
\end{equation}
where $T_{\textsc{b}}=\beta^{-1}$ is the bath's initial temperature.

\end{document}